\documentclass[a4paper,11pt]{scrartcl}

\usepackage{fullpage}
\usepackage{lineno,hyperref}

\usepackage{graphicx}
\usepackage{amsmath}
\usepackage{bm}

\usepackage{color}
\usepackage{tikz}
\usetikzlibrary{shapes,fit}
\usetikzlibrary{calc}
\usetikzlibrary{positioning}
\usetikzlibrary{decorations.pathreplacing}

\usepackage{pgfplots}
\usepackage{pgfplotstable}
\pgfplotsset{compat=1.9}

\usepackage{xifthen}
\usepackage{MnSymbol}

\usepackage{graphicx}

\usepackage{soul}

\usepackage[printonlyused]{acronym}
\usepackage{url}
\usepackage{colortbl}
\definecolor{VeryLightGray}{rgb}{0.9,0.9,0.9}
\definecolor{Color1}{rgb}{0.843, 0.098, 0.110} 
\definecolor{Color2}{rgb}{0.992, 0.682, 0.380} 
\definecolor{Color3}{rgb}{0.671, 0.851, 0.914} 
\definecolor{Color4}{rgb}{0.173, 0.482, 0.714} 
\newcommand{\minitab}[2][l]{\begin{tabular}{#1}#2\end{tabular}}
\usepackage{ctable}
\usepackage{multirow}
\usepackage{hhline}
\usepackage{eso-pic}
\usepackage{mathtools}
\usepackage{overpic}
\usepackage{changes}
\usepackage{authblk}

\begin{document}

\title{Throughput Optimizations for FPGA-based\\ Deep Neural Network Inference\thanks{This article has been published in the Journal \emph{Microprocessors and Microsystems} 60C (2018) pp. 151-161, \newline DOI: \texttt{10.1016/j.micpro.2018.04.004}\newline \textcopyright 2018. This manuscript version is made available under the CC-BY-NC-ND 4.0 license\newline \url{http://creativecommons.org/licenses/by-nc-nd/4.0/}}}

\author[1]{Thorbj\"orn Posewsky} 
\affil[1]{\large Institute of Embedded Systems,
Hamburg University of Technology (TUHH)  \protect\\
21073 Hamburg, Germany} 

\author[2]{Daniel Ziener}
\affil[2]{Computer Architecture for Embedded Systems,
University of Twente  \protect\\
7500 AE  Enschede, The Netherlands  \protect\\
Email: d.m.ziener@utwente.nl}


\date{ }
\maketitle

\begin{abstract}
Deep neural networks are an extremely successful and widely used technique for various pattern recognition and machine learning tasks.
Due to power and \mbox{resource} constraints, these computationally intensive networks are difficult to implement in embedded systems.
Yet, the number of applications that can \mbox{benefit} from the mentioned possibilities is rapidly rising.
In this \mbox{paper}, we \mbox{propose} novel architectures for the inference of previously learned and \mbox{arbitrary} deep neural networks on FPGA-based SoCs that are able to overcome these \mbox{limitations}.
Our key contributions include the reuse of previously transferred weight matrices across multiple input samples, which we refer to as batch \mbox{processing}, and the usage of compressed weight matrices, also known as \mbox{pruning}.
An \mbox{extensive} evaluation of these optimizations is presented. 
Both techniques \mbox{allow} a \mbox{significant} mitigation of data transfers and speed-up the network \mbox{inference} by one order of magnitude.
At the same time, we surpass the data throughput of fully-featured x86-based systems while only using a fraction of their energy \mbox{consumption}.

\end{abstract}


\section{Introduction and Motivation}

For more and more people, \emph{\aclp{DNN}} (\acsp{DNN}) have become a substantial part of their everyday life.
Applications like image classification \cite{CNNLIR} or speech recognition \cite{DBNLVCSR} are used by millions on their wearables, smartphones, or tablets. 
This applies not only to mobile computing, it also holds true for related areas like computer vision or robotics.
However, these emerging areas face restrictive power requirements and limited processing power, in contrast to high-performance computing, which is more often associated with deep learning techniques.

In order to achieve state-of-the-art and beyond classification rates in tasks like object recognition,
the number of artificial neurons and layers in \acsp{DNN} has grown to ever new records in the past years. 
Aside from a significantly increased demand for computational power, the size needed to store such networks has similarly increased.
For embedded devices, this is particularly challenging since memory is typically a scarce resource and, more importantly, the access to off-chip memories represents the dominating factor when considering the energy consumption \cite{DBLP:journals/corr/HanMD15}.
Hence, to lower both \acs{DNN} inference time and energy-consumption, this work focuses on techniques that reduce the amount of data to be transferred.

The first technique, called \emph{batch processing}, originates from applications that use or even require \emph{multiple} inferences of \acsp{DNN} with \emph{similar inputs} (also referred as \emph{samples}) before proceeding to the next step.
For example, the movement of UAVs, robots, or autonomous cars requires that images from different directions are evaluated before the next move is determined \cite{DRIVEPX}.
Deploying speech recognition at scale (i.e. in data centers) is another example where a study \cite{DBLP:journals/corr/AmodeiABCCCCCCD15} reports that a sequential processing of requests is inefficient due to the memory bound as well as a limited amount of exploitable parallelism.
\mbox{Instead}, \mbox{grouping} \mbox{multiple} samples together and processing this so-called batch can \mbox{often} \mbox{significantly} \mbox{increase} throughput in cases where several \acs{DNN} \mbox{inferences} are \mbox{necessary} or a small latency increase is tolerable.

The second technique investigated in this work, now known as \emph{pruning}, represents a form of \acs{DNN} compression \cite{lecun-90b} \cite{DBLP:journals/corr/HanMD15}. 
Instead of reusing data as in batch processing, pruning reduces the number of synaptic connections to other neurons such that the overall amount of data is reduced.
As described before, the tendency of growing \acsp{DNN} also increases the possibility of redundant connections.
Here, pruning can help eliminate these connections with minor, if any, accuracy drops for tasks such as classification.

While batch processing is a standard technique for an efficient \acs{DNN} training \cite{lecun-98b} (called mini-batch processing in the context of stochastic gradient descent), it is rarely used for the inference of \acsp{DNN}. In \cite{PZ16}, we showed 
how this concept affects the design of hardware accelerators for the \mbox{inference} of \acsp{DNN} (\emph{forward-propagation})  
and what latency consequences are imposed by realizing this concept in dedicated hardware.

Similarly, only a very limited number of previous works exists considering hardware-based support for pruned \acsp{DNN}. In this paper, we extend \cite{PZ16} and \cite{PZ18} and show
how a complete streaming architecture for arbitrarily pruned \acsp{DNN} can be designed as opposed to designs with partially or completely embedded parameters.

The contribution of this paper includes all the above mentioned aspects using an embedded \acs{FPGA}-based \acs{SoC} with limited external memory bandwidth.
Furthermore, we show for the first time an extensive evaluation and direct comparison of both techniques using the same \acs{DNN} networks and data sets. This includes for both techniques and designs expectable
\begin{itemize}
\item throughput gains,
\item accuracy variations, and
\item hardware design consequences and restrictions of complete \emph{streaming} architectures.
\end{itemize} 
We focus particularly on an efficient inference of \emph{fully-connected} \acsp{DNN} since these layers are the most memory-intensive and build the foundation for all of today's most successful network kinds. 

The rest of this paper is organized as follows:
Section~\ref{sec:rw} gives an overview of different network types, optimizations and corresponding hardware designs. Section~\ref{sec:background} provides some background information for neural network processing. 
The concepts and architectures of our accelerators are explained in Section~\ref{sec:concept} and~\ref{sec:arch}, respectively.
Section~\ref{sec:eval} continues with experimental results for different hardware configurations, software platforms, and using several network architectures. 
Finally, Section~\ref{sec:conc} concludes the work and highlights future research directions.

\section{Related Work}
\label{sec:rw}

In the past two decades, several hardware accelerators for various kinds of neural networks were introduced.
Many, and in particular early works, target shallow network architectures with few neurons or synaptic connections.
Two comprehensive studies that compare designs implemented on \acsp{FPGA} or as \acsp{ASIC} are given in \cite{Dias2004945} and \cite{Misra2010239}.
While these works serve the purposes of their time, today they are no longer applicable or optimized for networks of the deep learning era since the number of hardware neurons or connections is no longer sufficient.

An accelerator that addresses these deep networks is presented in \cite{Pietras6927383}.
It is based on an array of so called \emph{\aclp{NPU}} (\acsp{NPU}) that are used to compute the majority of involved operations (e.g., vector-matrix operations) in parallel.
Although this approach uses similar to our batch processing design a \emph{\acl{TDM}} (\acs{TDM}) processing scheme with a fast hardware-based switch of layers, it only exploits parallelism for one sample and relies on an ethernet connection for the transfer of network stimuli.
This requires a very time consuming retransfer of the required weight matrices for every sample and is not directly deployable for mobile devices.

Recently, many accelerator designs for \emph{\aclp{CNN}} (\acsp{CNN}) were introduced.
\acsp{CNN} are often found in image and video recognition systems and typically use a series of kernels or convolution matrices prior to the above mentioned fully-connected network architecture \cite{DBLP:journals/corr/Schmidhuber14}.
Since the number of parameters for convolution matrices is typically only a fraction of the weights of fully-connected network layers, the exploitable compute parallelism is usually greater and thus favors hardware accelerators.
A typical design that addresses theses networks is called \emph{NeuFlow} and proposed in \cite{farabet-ecvw-11} and \cite{farabet-suml-11}.
It relies on a two-dimensional grid of \emph{\aclp{PT}} (\acsp{PT}) 
instead of a one-dimensional array of \acsp{NPU}.
This resembles the concept of a systolic array, but both the routes of the dataflow and the operation of the individual \acsp{PT} are reconfigurable.
However, as reported in \cite{nn-X}, the proposed design has scalability issues which is problematic for batch processing as shown in Section~\ref{sec:eval}.
As a consequence, a \acs{CNN} design with a linear array of processing elements (called collections) is shown in
\cite{jin2014efficient} and \cite{nn-X}, respectively.
Nonetheless, both designs are particularly designed to accelerate \acsp{CNN}. Internal buffer and routing elements for an efficient execution of multiple samples in fully-connected layers are missing.

A third important type of networks is known as \emph{\acl{RNN}} (\acs{RNN}) \cite{DBLP:journals/corr/Schmidhuber14}.
\acsp{RNN} allow the processing of input sequences through cyclical connections in the network architecture.
Like fully-connected layers, these networks are typically memory bound and thus make a parallel execution more difficult.
Consequently, corresponding designs are less frequent.
However, an early approach for a state-of-the-art \acs{RNN}, called LSTMs, using the same \acs{FPGA} as in this work is shown in \cite{chang2015recurrent}.

The theoretical foundation for our second accelerator with support for pruned neural networks was introduced by LeCun et al. in \cite{lecun-90b}.
Originally, it was used to improve generalization and speed of learning in shallow network architectures.
However, Han et al. \cite{DBLP:journals/corr/HanMD15} recently revived the technique for \acsp{DNN} and were able to reduce the number of connections by a factor between 9x and 13x.
Although the pruned networks included both convolutional and fully-connected layers, most connections could be removed in the memory-intensive fully-connected layers.
Furthermore, they also introduced a form of parameter quantization and a subsequent Huffman encoding for the pruned and quantized networks. 
A corresponding \acs{ASIC} design with large on-chip memories for the remaining parameters after pruning and quantization (without Huffamn encoding) is given in \cite{DBLP:journals/corr/HanLMPPHD16}.
As discussed later, our accelerator utilizes a similar format, presented in \cite{vuduc2003automatic}, for the resulting sparse matrices (e.g., after pruning) but does not embed parameters for specific \acsp{DNN} on-chip.
Instead, we propose a streaming architecture for arbitrary \acsp{DNN}.
Very recently their approach was further extended to support LSTMs for speech recognition on high-performance \acsp{FPGA} \cite{DBLP:journals/corr/HanKMHLLXLYWYD16}.
Compared to our design, they use more complex \acsp{DNN} for specific tasks and directly map them onto large \acsp{FPGA} (up to 10x larger than the one used in this work). 
Instead, our design focuses on embedded \acs{FPGA}-based \acsp{SoC} with very limited on-chip memory resources and much slower memory interconnects.
We specifically design interfaces between the \acs{SoC}'s processors, off-chip memory and \acs{FPGA} in order to optimize \acsp{DNN} on such low-end and low-power devices.

\section{Background}
\label{sec:background}

\begin{figure}[t]
\centering
\def\layersep{2.5cm}

\resizebox{!}{4cm}{
\begin{tikzpicture}[shorten >=1pt,->,draw=black, node distance=\layersep]
    \tikzstyle{every pin edge}=[<-,shorten <=1pt,draw=black]
    \tikzstyle{neuron}=[circle,fill=black!50,draw=black,minimum size=17pt,inner sep=0pt]
    \tikzstyle{input neuron}=[neuron, fill=green!20]; 
    \tikzstyle{output neuron}=[neuron, fill=red!20];  
    \tikzstyle{hidden neuron}=[neuron, fill=blue!20]; 
    \tikzstyle{empty neuron}=[neuron, fill=none, draw=none];
    \tikzstyle{annot} = [text width=4em, text centered]
    
    \pgfmathsetmacro{\numInputNeurons}{3}
    \pgfmathsetmacro{\numHiddenNeurons}{4}
    \pgfmathsetmacro{\numHiddenTwoNeurons}{4}
    \pgfmathsetmacro{\numOutputNeurons}{3}

    \foreach \name / \y in {1,...,\numInputNeurons}{
        \ifthenelse{\NOT 2 = \y}{
            \node[input neuron] (I-\name) at (0,-\y) {};
        }{
            \node[empty neuron] (I-\name) at (0,-\y) {\vdots};
        }
    }

    \foreach \name / \y in {1,...,\numHiddenNeurons}{
        \ifthenelse{\NOT 3 = \name}{
            \path [yshift=0.5cm] node[hidden neuron] (H-\name) at (\layersep,-\y cm) {};
        }{
            \path [yshift=0.5cm] node[empty neuron] (H-\name) at (\layersep,-\y cm) {\vdots};
        }
    }
    
    \foreach \name / \y in {1,...,\numHiddenTwoNeurons}{
        \ifthenelse{\NOT 3 = \name}{
            \path [yshift=0.5cm] node[hidden neuron] (H2-\name) at (2*\layersep,-\y cm) {};
        }{
            \path [yshift=0.5cm] node[empty neuron] (H2-\name) at (2*\layersep,-\y cm) {\vdots};
        }
    }

    \foreach \name / \y in {1,...,\numOutputNeurons}{
        \node[output neuron] (O-\name) at (3*\layersep,-\y cm) {};
    }

    \foreach \source in {1,...,\numInputNeurons}
        \foreach \dest in {1,...,\numHiddenNeurons}{
            \ifthenelse{\NOT 2 = \source \AND \NOT \dest = 3}{
                \path ([xshift=0.05cm]I-\source.east) edge (H-\dest);
            }{}
        }

    \foreach \source in {1,...,\numHiddenNeurons}
        \foreach \dest in {1,...,\numHiddenTwoNeurons}{
            \ifthenelse{\NOT 3 = \source \AND \NOT \dest = 3}{
                \path ([xshift=0.05cm]H-\source.east) edge (H2-\dest);
            }{}
        }

    \foreach \source in {1,...,\numHiddenTwoNeurons}
        \foreach \dest in {1,...,\numOutputNeurons}{
            \ifthenelse{\NOT 3 = \source}{
                \path ([xshift=0.05cm]H2-\source) edge (O-\dest);
            }{}
        }

    \node[annot,above of =H-1, node distance=1.0cm] (hl)  {Hidden Layer};
    \node[annot,right of = hl] (hl2) {Hidden Layer};
    \node[annot,left of  = hl]        {Input Layer};
    \node[annot,right of =hl2]        {Output Layer};
\end{tikzpicture}
}
\caption{Example neural network with two hidden layers ($L=4$). In this case, the output layer $j=4$ contains three neurons, whereas all previous layers $j=1 \dots 3$ contain an arbitrary number of neurons.}
\label{fig:ExampleDNN}
\end{figure}
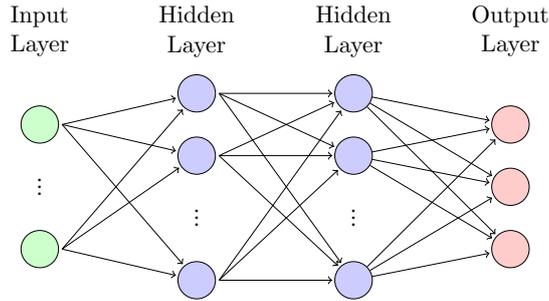

A typical neural network contains several layers $j = 1 \dots L$, where $L$ denotes the number of layers. A layer $j$ itself consists of $s_j$ neurons.
As already mentioned, one major goal of this work is to accelerate the processing of fully-connected layers in \acsp{DNN}.
These layers are characterized by a bipartite graph of neuron connections between two adjacent layers $j$ and $j+1$ for $1 \leq j \leq L-1$. 
For the rest of this work, we will specify the architecture of these networks through the number of neurons $s_j$ in each layer.
For example, a network with $L = 3$ layers is denoted by $s_0 \times s_1 \times s_2$.
The synaptic strength of a connection is modeled through a scalar value $w_{i,k}^{(j)}$ called \emph{weight} that represents the connection to the $i$-th neuron in layer $j+1$ from the $k$-th neuron in layer $j$.
A transition from layer $j$ to the next layer $j+1$ involves a \emph{weight matrix} $W^{(j)}$ where $w_{i,k}^{(j)}$ are the components.
The number of rows in $W^{(j)}$ equals the number of neurons $s_{j+1}$ in layer $j+1$ and the number of columns corresponds to the number of neurons $s_{j}$ in layer $j$.  Figure~\ref{fig:ExampleDNN} gives an example of a neural network with four layers.

The result of each neuron is computed by the following two functions:
First, the \emph{transfer function} which is a series of \emph{multiply-accumulate} (\acs{MAC}) operations of the outputs $a_k^{(j)}$ of connecting neurons in the layer $j$ and their corresponding weights $w_{i,k}^{(j)}$:
\[
z_i^{(j+1)} = \sum_{k=0}^{s_{j}} w_{i,k}^{(j)} \cdot a_k^{(j)}
\]
Second, a subsequent application of a non-linear function, called \emph{activation function} $\varphi$, with the result of the transfer function as argument:
\[
a_i^{(j+1)} = \varphi(z_i^{(j+1)})
\]
The outputs of this function are also referred to as activations for the sake of brevity.
A variety of different types of activation functions $\varphi$ are known in neural network literature.
For example, while before the deep learning era the so called \emph{sigmoid} function was found most frequently, today's most successful implementations usually deploy \emph{\aclp{ReLU}} (\acs{ReLU}) \cite{nair2010rectified} or variations of it \cite{DBLP:journals/corr/ClevertUH15}.
It is also not uncommon to utilize different functions in the same neural network, e.g., the sigmoid function for the output layer and \acsp{ReLU} for all other layers.
In order to ensure the application of our accelerator for various trained networks, the accelerator is able to choose between different functions at runtime.

\section{Concept}
\label{sec:concept}

On the hardware side, modern \acsp{FPGA} typically offer a rich set of \acs{DSP} and \acs{RAM} resources within their fabric that can be used to process these networks.
However, compared to the depth and layer size of deep neural networks, these resources are no longer sufficient for a full and direct mapping the way it was often done in previous generations of neural network accelerators.
For example, consider a network with $L=7$ layers and architecture $784 \times 2500 \times 2000 \times 1500 \times 1000 \times 500 \times 10$ that was proposed in \cite{DBLP:journals/corr/abs-1003-0358}.
The number of neurons is 8294 and the total size of the network weights is approximately 22 MB if each weight is encoded using 16 bits.
Compared to \acs{FPGA} platforms like the Zynq, where even the largest device is limited to 2020 \acs{DSP} slices and a total \acs{BRAM} size of less than 3 MB \cite[pp. 622]{zynqrefman}, a complete mapping with all neurons and weights directly onto the \acs{FPGA} is no longer possible. Here, new algorithms and architectural adaptions are required in order to enable the inference of \acsp{DNN} on platforms with such limited resources.

Modern and deep neural networks are usually partitioned into smaller \emph{sections} in order to process them on embedded \acsp{FPGA} platforms.
We refer to a \emph{section} as a certain number $m$ of neurons in a given layer $j$ with $m \leq s_{j+1}$ that can be processed in parallel through our hardware coprocessor with $m$ individual \emph{processing units}.
Each processing unit is responsible for the transfer function of exactly one neuron in each section.
By applying a time division multiplexing scheme, the whole network can be processed on these $m$ processing units and a subsequent 
activation function. Each \emph{processing unit} may consists of $r$ different computation resources, e.g., multipliers which are able to consume $r$ weights as inputs in parallel for the calculation of the transfer function. 
The number of processing units $m$ and the corresponding number of compute resources per processing unit $r$ indicate the degree of parallelism and depends on the number of available compute resources in hardware.
Since the network is fully-connected, the computation of layer $j$ requires that all previous layers $1 \dots j-1$ are \emph{completely} processed.
Consequently, a hardware implementation can only use parallelism in the current layer $j$ and not across multiple layers.

Due to the fact that the on-chip memory is not sufficient for storing all needed weights for an arbitrary layer, only the weights for processing the current section can be loaded from external memory.
When comparing the size of the input data ($s_{j}$ values), the output data ($m$ values), and in particular the weights ($\approx s_{j} \times m$ values), it can be seen that the transfer of the weight matrix is very costly.
Three concepts for reducing memory data transfers are discussed in the following: Weight encoding in order to reduce the used weight bits, batch processing to reuse weights which are already on-chip \cite{PZ16}, and pruning to remove weights such that it is unnecessary to transfer them. 

\subsection{Weight Encoding}

An enormous impact on the throughput, complexity, and amount of needed memory resources has the encoding of the weight matrices $W^{(j)}$ with the corresponding individual weights $w_{i,k}^{(j)}$. Software-based implementations often use floating point weights, whereas the most hardware implementations using fixed point representations. On hardware implementations, the encoding format can be freely chosen.
Accuracy evaluations show that often the accuracy loss through weight bit reduction is negligible compared to the advantages of  increased weight memory throughput and reduced operation complexity  \cite{farabet-ecvw-11,nn-X,Chen:2014:DSH:2541940.2541967}. Extreme approaches reducing the weights to only a single bit which is called binary neural networks (BNNs) processing \cite{DBLP:journals/corr/UmurogluFGBLJV16}. However, the accuracy of such BNNs  is relatively low compared to other approaches. Reducing the used weight bits has mainly the advantage of reduced data amount for storing and transferring weights. Furthermore, the computation complexity might be reduced, e.g., transforming the multiplication into an addition on the BNN approach. However, due to the often usage of hardware multipliers or DSP resources with fixed input width, only a small reduction of used resources is achievable. Most hardware implementations are able to perform every clock cycle one operation, i.e. multiplication. Therefore, the reduction of the used weight bits do not increase the overall  throughput of the processing elements, if the weight transfer is ignored.

\subsection{Batch Processing}

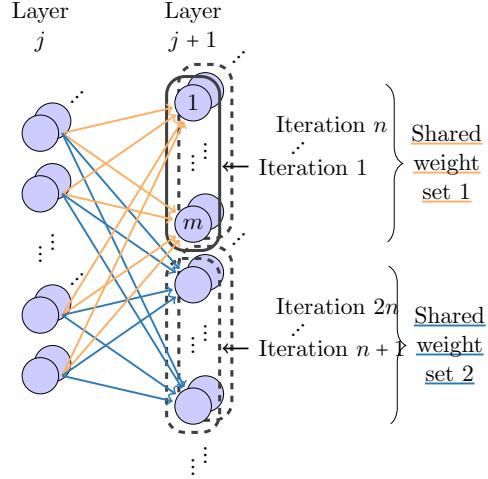
\begin{figure}[t]
\centering
\def\layersep{2.5cm}

\pgfdeclarelayer{sampleone layer}
\pgfdeclarelayer{sampletwo layer}
\pgfsetlayers{sampletwo layer,sampleone layer,main} 
\resizebox{!}{6.5cm}{%
\begin{tikzpicture}[shorten >=1pt,->,draw=black!80, node distance=\layersep]
    \tikzstyle{every pin edge}=[<-,shorten <=1pt,draw=black]
    \tikzstyle{neuron}=[circle,fill=black!25,draw=black,minimum size=17pt,inner sep=0pt]
    \tikzstyle{input neuron}=[neuron, fill=blue!20];
    \tikzstyle{hidden neuron}=[neuron, fill=blue!20];
    \tikzstyle{empty neuron}=[neuron, fill=none, draw=none];
    \tikzstyle{annot}=[text width=4em, text centered]
    
    \pgfmathsetmacro{\numInputNeurons}{5}
    \pgfmathsetmacro{\numHiddenNeurons}{6}
    \pgfmathsetmacro{\lastHiddenNeuron}{7} 
    \pgfmathsetmacro{\shiftAmountOne}{0.2}
    \pgfmathsetmacro{\shiftAmountTwoInput}{0.6}
    \pgfmathsetmacro{\shiftAmountTwoHidden}{0.75}
    \definecolor{weightset1}{rgb}{0.992, 0.682, 0.380} 
    \definecolor{weightset2}{rgb}{0.173, 0.482, 0.714} 
    \def\weightsetthickness{0.9pt}
    \newcommand{\IterationName}{Iteration }

    \foreach \name / \y in {1,...,\numInputNeurons}{
        \ifthenelse{\NOT 3 = \y}{
            \node[input neuron] (I-\name) at (0,-\y) {};
        }{
            \node[empty neuron] (I-\name) at (0,-\y) {\vdots};
        }
    }

    \foreach \name / \y in {1,...,\numHiddenNeurons}{
        \ifthenelse{\NOT 2 = \name \AND \NOT 5 = \name}{
            \ifthenelse{1 = \name}{
                \path [yshift=0.5cm] node[hidden neuron] (H-\name) at (\layersep,-\y cm) {$1$};
            }{
                \ifthenelse{3 = \name}{
                    \path [yshift=0.5cm] node[hidden neuron] (H-\name) at (\layersep,-\y cm) {$m$};
                }{
                    \path [yshift=0.5cm] node[hidden neuron] (H-\name) at (\layersep,-\y cm) {};
                }
            }
        }{
            \path [yshift=0.5cm] node[empty neuron] (H-\name) at (\layersep,-\y cm) {\vdots};
        }
    }
    \path [yshift=0.5cm] node[empty neuron] (H-\lastHiddenNeuron) at (\layersep,-\lastHiddenNeuron cm) {\vdots};
            
    \node[rectangle,rounded corners=0.3cm,draw=black!75,line width=0.5mm, fit=(H-1) (H-3)] (batch1) {};
    \node[xshift=-0.15cm,yshift=-0.05cm,pin={[pin edge={black,thick}]right:{\IterationName $1$} }] at (batch1.east) {};
    \node[rectangle,rounded corners=0.3cm,draw=black!75,dashed,line width=0.5mm, fit=(H-4) (H-6)] (batchn1) {};
    \node[xshift=-0.15cm,yshift=-0.05cm,pin={[pin edge={black,thick}]right:{\IterationName $n+1$} }] at (batchn1.east) {};


    \foreach \source in {1,...,\numInputNeurons}
        \foreach \dest in {1,...,\numHiddenNeurons}{
            \ifthenelse{\NOT 3 = \source \AND \NOT 2 = \dest \AND \dest < 4}{
                \path ([xshift=0.05cm]I-\source.east) edge[draw=weightset1,line width=\weightsetthickness] (H-\dest);
            }{}
            \ifthenelse{\NOT 3 = \source \AND \dest > 3 \AND \NOT \dest = 5}{
                \path ([xshift=0.05cm]I-\source.east) edge[draw=weightset2,line width=\weightsetthickness] (H-\dest);
            }{}
        }


    \node[annot,above of=H-1, node distance=1.25cm] (hl) {Layer\qquad $j+1$};
    \node[annot,left of=hl] {Layer\qquad $j$};

    \begin{pgfonlayer}{sampleone layer}
      \foreach \name / \y in {1,...,\numInputNeurons}{
          \ifthenelse{\NOT 3 = \y}{
              \node[input neuron, shift={(\shiftAmountOne,\shiftAmountOne)}] (I-\name) at (0,-\y) {};
          }{
              \node[empty neuron, shift={(\shiftAmountOne,\shiftAmountOne)}] (I-\name) at (0,-\y) {\vdots};
          }
      }
  
      \foreach \name / \y in {1,...,\numHiddenNeurons}{
          \ifthenelse{\NOT 2 = \name \AND \NOT 5 = \name}{
              \path [yshift=0.5cm] node[hidden neuron, shift={(\shiftAmountOne,\shiftAmountOne)}] (H2-\name) at (\layersep,-\y cm) {};
          }{
              \path [yshift=0.5cm] node[empty neuron, shift={(\shiftAmountOne,\shiftAmountOne)}] (H2-\name) at (\layersep,-\y cm) {\vdots};
          }
      }
      \path [yshift=0.5cm] node[empty neuron, shift={(\shiftAmountOne,\shiftAmountOne)}] (H-\lastHiddenNeuron) at (\layersep,-\lastHiddenNeuron cm) {\vdots};
      
      \node[rectangle,rounded corners=0.3cm,draw=none, fit=(H2-1) (H2-3)] (batch2) {};
      \node[yshift=0.05cm,xshift=1.08cm] at (batch2.east) {$\udots$};
      \node[rectangle,rounded corners=0.3cm,draw=none, fit=(H2-4) (H2-6)] (batch2) {};
      \node[yshift=0.05cm,xshift=1.08cm] at (batch2.east) {$\udots$};
      
      \node[rectangle,rounded corners=0.3cm,dashed,draw=black!75,line width=0.5mm, fit=(H2-1) (H2-3)] (batch2) {};
      \node[rectangle,rounded corners=0.3cm,draw=black!75,dashed,line width=0.5mm, fit=(H2-4) (H2-6)] (batch2) {};
    \end{pgfonlayer}
    
    \begin{pgfonlayer}{sampletwo layer}
      \foreach \name / \y in {1,...,\numInputNeurons}{
          \ifthenelse{\y = 1 \OR \y = 4}{
              \node[empty neuron, shift={(\shiftAmountTwoInput,\shiftAmountTwoInput)}] (I-\name) at (0,-\y) {$\udots$};
          }{}
      }
  
      \foreach \name / \y in {1,...,\numHiddenNeurons}{
          \ifthenelse{1 = \name \OR 4 = \name}{
              \path [yshift=0.5cm]
              node[empty neuron, shift={(\shiftAmountTwoHidden,\shiftAmountTwoHidden)}] (H3-\name) at (\layersep,-\y cm) {$\udots$};
          }{
              \path [yshift=0.5cm]
              node[empty neuron, shift={(\shiftAmountTwoHidden,\shiftAmountTwoHidden)}] (H3-\name) at (\layersep,-\y cm) {};
          }
      }
      
      \node[draw=none, fit=(H3-1) (H3-3)] (batchn) {};
      \node[xshift=-0.6cm,yshift=-0.1cm,pin={[pin edge={white,ultra thick}]right:{\IterationName $n$} } ] at (batchn.east) {};
      \node[draw=none, fit=(H3-4) (H3-6)] (batch2n) {};
      \node[xshift=-0.6cm,yshift=-0.1cm,pin={[pin edge={white,ultra thick}]right:{\IterationName $2n$} } ] at (batch2n.east) {};
      
      \setulcolor{weightset1}
      \setul{\weightsetthickness}{\weightsetthickness}
      \draw [-,decorate,decoration={brace,amplitude=8pt,raise=3.15cm},xshift=0pt,yshift=0pt]
      (H-1.north) -- (H-3.south) 
      node [black,midway,xshift=4.15cm,align=center]{
      \ul{Shared}\\\ul{weight}\\\ul{set 1}
      };
      
      \setulcolor{weightset2}
      \setul{\weightsetthickness}{\weightsetthickness}
      \draw [-,decorate,decoration={brace,amplitude=8pt,raise=3.2cm},xshift=0pt,yshift=0pt]
      (H-4.north) -- (H-6.south) 
      node [black,midway,xshift=4.2cm,align=center] {
      \ul{Shared}\\\ul{weight}\\\ul{set 2}
      };
    \end{pgfonlayer}
\end{tikzpicture}
}
\caption{Conceptual batch processing with batch size $n$ and section size $m$.
All $m$ neurons in a section are processed in parallel.
The first section of all $n$ samples shares the same collection of weights.
The second section of all $n$ samples shares the next collection of weights, and so on.}
\label{fig:BatchProcessing}
\end{figure}

A straightforward section by section processing for just one sample has the drawback of exchanging the costly transferred weights for every new section.
This huge demand of memory transfers turns the interface to the external memory into the major bottleneck for fully-connected layer processing. The input data $a_k^{(j)}$ (i.e., results of the previous layer), however, is needed for all sections in the current layer.
Therefore, it should be cached in on-chip memories during the complete processing.
The main contribution of our batch processing approach is the reuse of already transferred and stored weights for one section by processing different input samples through time division multiplexing.
This processing scheme is visualized in Figure~\ref{fig:BatchProcessing}.

Given a batch of $n$ different input samples, the algorithm starts by processing all first sections of the $n$ samples before proceeding to all second sections of the $n$ samples.
Thereby, all iterations $1 \dots n$ use the same set of weights, however, distinct input samples.
Only before the processing of the second sections, a new set of weights is transferred.
This technique reduces the amount of memory transfers significantly and, therefore, mitigates the memory interface bottleneck.
Note that for general matrix operations, similar processing schemes were already  discussed in earlier works \cite{McKellar:1969:OMM:362875.362879}.
However, as shown in Section~\ref{sec:arch}, our design specifically incorporates all \acsp{DNN} operations and allows an interleaving of this concept and subsequent operations (i.e., activation functions) in order to further enhance the throughput.

\subsection{Pruning}\label{sec:con:pruning}
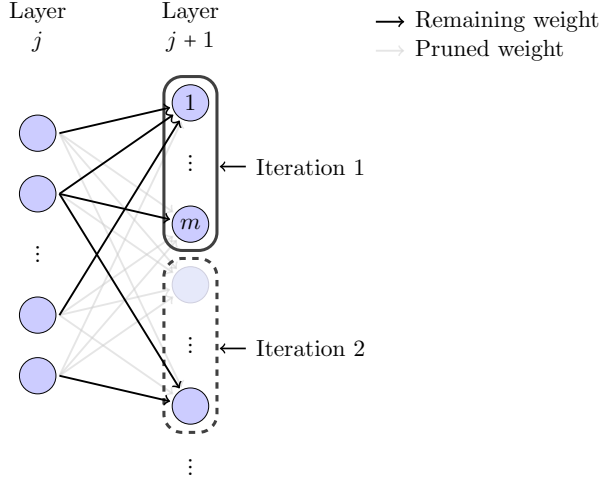
\begin{figure}[t]
\centering
\def\layersep{2.5cm}

\pgfdeclarelayer{back}
\pgfsetlayers{back,main}

\makeatletter
\pgfkeys{%
  /tikz/on layer/.code={
    \def\tikz@path@do@at@end{\endpgfonlayer\endgroup\tikz@path@do@at@end}%
    \pgfonlayer{#1}\begingroup%
  }%
}
\makeatother

\resizebox{!}{6.5cm}{%
\begin{tikzpicture}[shorten >=1pt,->,draw=black!80, node distance=\layersep]
    \tikzstyle{every pin edge}=[<-,shorten <=1pt,draw=black]
    \tikzstyle{neuron}=[circle,fill=black!25,draw=black,minimum size=17pt,inner sep=0pt]
    \tikzstyle{input neuron}=[neuron, fill=blue!20];
    \tikzstyle{hidden neuron}=[neuron, fill=blue!20];
    \tikzstyle{empty neuron}=[neuron, fill=none, draw=none];
    \tikzstyle{pruned neuron}=[neuron, fill=red!20!green!20!blue!20, draw=red!50!green!50!blue!50, opacity=.5];
    \tikzstyle{annot}=[text width=4em, text centered]
    
    \pgfmathsetmacro{\numInputNeurons}{5}
    \pgfmathsetmacro{\numHiddenNeurons}{6}
    \pgfmathsetmacro{\lastHiddenNeuron}{7} 
    \pgfmathsetmacro{\shiftAmountOne}{0.2}
    \pgfmathsetmacro{\shiftAmountTwoInput}{0.6}
    \pgfmathsetmacro{\shiftAmountTwoHidden}{0.75}
    \definecolor{weightset1}{rgb}{0.0,0.0,0.0}
    \definecolor{weightset2}{rgb}{0.8,0.8,0.8}
    \def\weightsetthickness{0.9pt}
    \newcommand{\IterationName}{Iteration }

    \foreach \name / \y in {1,...,\numInputNeurons}{
        \ifthenelse{\NOT 3 = \y}{
            \node[input neuron] (I-\name) at (0,-\y) {};
        }{
            \node[empty neuron] (I-\name) at (0,-\y) {\vdots};
        }
    }

    \foreach \name / \y in {1,...,\numHiddenNeurons}{
        \ifthenelse{\NOT 2 = \name \AND \NOT 4 = \name \AND \NOT 5 = \name}{
            \ifthenelse{1 = \name}{
                \path [yshift=0.5cm] node[hidden neuron] (H-\name) at (\layersep,-\y cm) {$1$};
            }{
                \ifthenelse{3 = \name}{
                    \path [yshift=0.5cm] node[hidden neuron] (H-\name) at (\layersep,-\y cm) {$m$};
                }{
                    \path [yshift=0.5cm] node[hidden neuron] (H-\name) at (\layersep,-\y cm) {};
                }
            }
        }{
            \ifthenelse{4 = \name}{
                \path [yshift=0.5cm] node[pruned neuron] (H-\name) at (\layersep,-\y cm) {};
            }{
                \path [yshift=0.5cm] node[empty neuron] (H-\name) at (\layersep,-\y cm) {\vdots};
            }
        }
    }
    \path [yshift=0.5cm] node[empty neuron] (H-\lastHiddenNeuron) at (\layersep,-\lastHiddenNeuron cm) {\vdots};
            
    \node[rectangle,rounded corners=0.3cm,draw=black!75,line width=0.5mm, fit=(H-1) (H-3)] (batch1) {};
    \node[xshift=-0.15cm,yshift=-0.05cm,pin={[pin edge={black,thick}]right:{\IterationName $1$} }] at (batch1.east) {};
    \node[rectangle,rounded corners=0.3cm,draw=black!75,dashed,line width=0.5mm, fit=(H-4) (H-6)] (batchn1) {};
    \node[xshift=-0.15cm,yshift=-0.05cm,pin={[pin edge={black,thick}]right:{\IterationName $2$} }] at (batchn1.east) {};


    \foreach \source in {1,...,\numInputNeurons}
        \foreach \dest in {1,...,\numHiddenNeurons}{
            \ifthenelse{\NOT 3 = \source \AND 
                        \NOT 2 = \dest \AND   
                        \NOT 5 = \dest \AND   
                        \NOT 4 = \dest        
                       }
            {
                
                \ifthenelse{
                    \(1 = \dest \AND 1 = \source\) \OR
                    \(1 = \dest \AND 2 = \source\) \OR
                    \(3 = \dest \AND 2 = \source\) \OR
                    \(1 = \dest \AND 4 = \source\) \OR
                    \(6 = \dest \AND 2 = \source\) \OR
                    \(6 = \dest \AND 5 = \source\)
                }
                {
                    \path ([xshift=0.05cm]I-\source.east) edge[draw=weightset1,line width=\weightsetthickness] (H-\dest);
                }{
                    \path ([xshift=0.05cm]I-\source.east) edge[draw=weightset2,opacity=.5,on layer=back,line width=\weightsetthickness] (H-\dest);
                }
            }{
                \ifthenelse{\NOT 3 = \source \AND 4 = \dest}{
                    \path ([xshift=0.05cm]I-\source.east) edge[draw=weightset2,opacity=.5,on layer=back,line width=\weightsetthickness] (H-\dest);
                }{}
            }
        }


    \node[annot,above of=H-1, node distance=1.25cm] (hl) {Layer\qquad $j+1$};
    \node[annot,left of=hl] {Layer\qquad $j$};
    
    \node[draw=none,thick,rounded corners=2pt,below left=2mm,right of=hl,node distance=5cm] {%
    \begin{tabular}{@{}r@{ }l@{}}
        \raisebox{2pt}{\tikz{\draw[draw=weightset1,line width=\weightsetthickness] (0,0) -- (5mm,0);}} & Remaining weight\\
        \raisebox{2pt}{\tikz{\draw[weightset2,opacity=.5,on layer=back,line width=\weightsetthickness] (0,0) -- (5mm,0);}} & Pruned weight
    \end{tabular}};
\end{tikzpicture}
}
\caption{Example of a pruned \acs{DNN} layer.
Up to $m$ neurons in a section can be processed in parallel.
Computations are only required for the remaining weights and can be entirely skipped for neurons with only pruned weights.
}
\label{fig:PruningConcept}
\end{figure}

In order to reduce the amount of data to transfer from the memory and for calculation, it is possible to remove some connections entirely.
After some initial iterations of the training phase, small weights which are below a certain threshold $\delta$ can be set to zero:
\[
w_{i,k}^{(j)} < \delta  \qquad  \xRightarrow[\text{iterations}]{\text{following}}  \qquad  w_{i,k}^{(j)} := 0
\]
Subsequently, these pruned weights are kept at zero and the remaining weights are refined in the following iterations of the training phase.
While this can potentially reduce the accuracy if too many weights are pruned, it was shown that over 90\% of the weights in fully-connected layers of common \acsp{CNN} can be pruned without noticeable accuracy drops \cite{DBLP:journals/corr/HanMD15}.
An example of a pruned layer is shown in Figure~\ref{fig:PruningConcept}.

Since weights with the value zero neither influence the result of the transfer nor the result of the activation function, these weights don't have to be stored in memory, transferred, and used for computations.
However, by pruning weights, the weight matrix becomes sparse and the hardware needs to be designed in a way that the involved calculations are computed efficiently.
Additionally, this presupposes a suitable format to store the sparse weight matrices in a continuous way with the smallest possible footprint.
Details about sparse matrix computation and storage are further discussed in Section~\ref{sec:arch}.

\subsection{Throughput Discussion}
\label{sec:con:throughput}

Assuming that the computation resources of a general hardware architecture are able to process every clock cycle one value, the following amount of clock cycles are needed for the computation of layer $j+1$ with $s_{j+1}$ neurons and $s_j$ input activations for a total of $N$ input samples:

\[
\Bigl\lceil \frac{s_{j+1}}{m} \Bigr\rceil \cdot \Bigl\lceil  \frac{s_j \cdot (1 - q_\text{prune}^{(j)})}{r} \Bigr\rceil  \cdot N
\]

whereas $m$ is the number of neurons which can be processed in parallel, and $r$ is number of parallel processed operations per neuron. The pruning factor $0 \leq q_\text{prune}^{(j)} \leq 1$ expresses the reduction of weights by pruning. For example, if 90\% of all weights $w_{i,k}^{(j)}$ are pruned, then the pruning factor is $q_\text{prune}^{(j)} = 0.9$.  Through elaborated pipelining, most hardware architectures achieve a throughput of one value per computation resource, just as well as our architectures, presented in Section \ref{sec:arch}. 
For large $s_{j+1}$, $s_j$, and $N$, we can calculate the approximated  processing time:

\[
t_{calc} \approx \frac{ s_{j+1} \cdot s_j \cdot N \cdot (1 - q_\text{prune}^{(j)})}{m \cdot r \cdot f_{pu}},
\]

whereas   $f_{pu}$ is the clock frequency of the processing units.

However, this approximation does not consider the transfer time of the weight matrix $w^{(j)}$ from the external memory.
The time to transfer all weights for the calculation of layer $j+1$ for $N \gg n$ input samples is

\[
t_{\text{mem}}= \frac{s_{j+1} \cdot s_j \cdot b_\text{weight} \cdot N }{T_\text{mem} \cdot n},
\]

where $n$ is the batch size, $b_\text{weight}$ is the size of each weight, and $T_\text{mem}$ the actual memory throughput. If weight pruning is used, the number of weight $s_{j+1} \cdot s_j$  is reduced by the pruning factor $q_\text{prune}^{(j)}$. However, additional information must be stored in the memory to determine the positions of the remaining weights (see Section~\ref{sec:arch:pruning}). Therefore, the size for storing a $b_\text{weight}$ bit weight is increased by the factor $q_\text{overhead} \geq 1$. The resulting formula with pruning is:

\[
t_{\text{mem}}= \frac{s_{j+1} \cdot s_j \cdot b_\text{weight} \cdot q_\text{overhead} \cdot (1 - q_\text{prune}^{(j)}) \cdot N}{T_\text{mem} \cdot n}
\]

The output calculation and the weight transfers are running in parallel. Therefore, the maximum of both times determines the overall processing time $t_\text{proc}$:

\[
t_\text{proc} = \text{max}(t_\text{calc}, t_{\text{mem}})
\]

It can be seen that pruning is a very efficient measure to increase the throughput of embedded neural network processing due to the fact that the weight transfers are reduced as well as the number of calculations. However, an accuracy reduction might be taken into account.
On the other hand, batch processing has no influence on the overall accuracy while significantly reducing the number of weight transfers. However, the number of operations remain the same, and an increased processing latency has to be taken into account.

By comparing the reduction of operations and memory transfers through pruning, it can be seen that the number of calculations is reduced by the inverse of the pruning factor $1 - q_\text{prune}^{(j)}$, whereas the number of data to transfer is only reduced by $ (1- q_\text{prune}^{(j)}) \cdot q_\text{overhead}$. In comparison, batch processing reduces only the amount of data transfers. Therefore, both methods are very effective in order to increase the overall throughput. 

The network architecture, the achieved pruning factor $q_\text{prune}^{(j)}$, and the size of each weight $b_\text{weight}$ are determined and optimized in the learning phase to achieve the required accuracy. To optimize the overall throughput of the hardware architecture with the given and above mentioned parameters,  the number of processing resources $m \cdot r$ can be maximized and an optimal batch size $n_{\text{opt}}$ can be calculated. This is achieved by setting $t_{mem} =  t_{calc}$,
i.e. neither the memory interface nor the MAC units have to wait for data or requests.
This optimal batch size $n_{opt}$ can be calculated with 

\[
n_{\text{opt}} \approx \frac{m \cdot r \cdot f_{pu} \cdot b_\text{weight} \cdot q_\text{overhead} }{T_\text{mem}}.
\]

\section{Architecture}
\label{sec:arch}

\begin{figure}[t]

\centering
\includegraphics[width=0.75\linewidth]{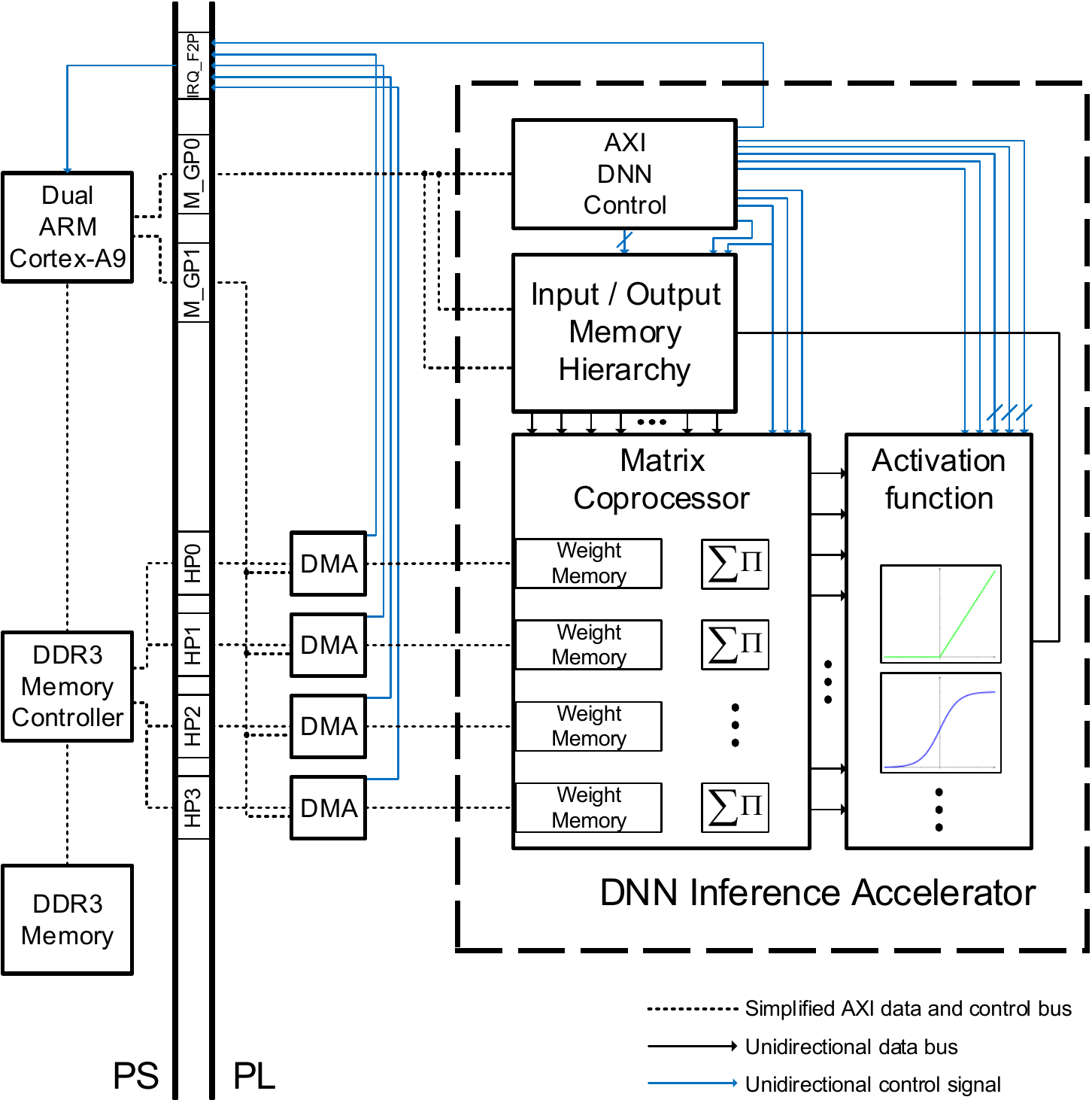}
\caption{Overview of our \acs{DNN} accelerator with the Zynq processing system (\acs{PS}) on the left and the custom accelerator inside the programmable logic (\acs{PL}) on the right. The connecting \acs{PS}-\acs{PL} interfaces are shown in between. In addition, four \acs{DMA} master peripherals are used for the weight transfer.}
\label{fig:BigPicture}  
\end{figure}

We have implemented two architectures to demonstrate the batch processing and pruning approach on Xilinx's \emph{Zynq-7000 All Programmable SoC} platform \cite{zynqrefman}.
This \acs{SoC} represents an affordable, low power device with a recent \acs{FPGA} fabric that is suitable for various embedded systems.
An overview visualizing the overall accelerator structure with generic components for both designs and all related Zynq peripherals is shown in Figure~\ref{fig:BigPicture}.

FPGA-based coprocessors for the Zynq usually depend highly on the interfaces between the \emph{processing system} (\acs{PS}) and the \emph{programmable logic} (\acs{PL}) in order to achieve the highest transfer bandwidth. 
In our case, this is especially true for the \acs{DDR}3 memory controller that resides inside the \acs{PS} and is used to retrieve the network weights.

All major connections that cross the boundary of our actual \acs{DNN} accelerator are indicated as dashed lines in Figure~\ref{fig:BigPicture}.
These buses pass several interconnects and controllers, both inside the \acs{PS} and the \acs{PL}, which are necessary for the communication but are omitted in the visualization in order to simplify the overview and to focus on the most important aspects.

In general, the software running on the ARM cores of the Zynq is used to configure and monitor both the control unit of the accelerator and all four \acs{DMA} engines.
It is also meant to transfer the network input and outputs.

The actual processing begins as soon as both the first inputs from the software and the first weights from a burst transfer of the \acs{DMA} engines arrive.
For this purpose, both accelerators share an overall similar architecture which is divided into four major \acsp{IP}.
However, depending on the concrete design, each of these \acsp{IP} is differently implemented.
Similarities and differences are detailed in the following:

\subsection{Control Unit}

The first \acs{IP}, called \emph{AXI \acs{DNN} Control}, is the control unit of the three remaining datapath \acsp{IP} in Figure~\ref{fig:BigPicture}.
In addition, it stores metadata, like the dimension of the matrix operation, or certain runtime adjustable settings, like the type of the activation function (e.g., \acs{ReLU} or sigmoid).
It also monitors the current processing stage and is able to precisely inform the software side about requests and events like required data transfers.
Furthermore, it stores additional design specific information (e.g., the batch size for the batch processing design).

\subsection{Input / Output Memory Hierarchy}

Both accelerators have an internal memory hierarchy that is used to store input and output activations for the currently calculated layer.
While the input for the first layer needs to be copied by the ARM cores, the inputs for the following layers are always outputs of previous layers and thus computed and stored inside the memory hierarchy. 
The flexibility to arbitrarily act as input or output requires a certain hierarchy of multiplexers, demultiplexers and multiple memory ports since the data must be accessible by the processing system and multiple compute units inside the programmable logic.
Depending on the design, some degree of redundancy is required in order to avoid pipeline stalls.
This is further explained in Section~\ref{sec:arch:pruning}.

\subsection{Matrix Coprocessor}
\label{sec:TF}

The \acs{IP} with the largest resource utilization is the matrix coprocessor that computes the transfer function, i.e., the weighted sum of inputs $z_i^{(j)}$.
This involves matrix-vector (pruning) or matrix-matrix (batch processing) operations that are mainly implemented with multiply-accumulate units (\acsp{MAC}) by using \acs{DSP} slices.
We use a fixed point data format, known as Q7.8, that consists of one sign bit, seven integer bits and eight fractional bits.
Although there exist first results that use fewer bits for both weights and activations (e.g., between $1$ and $8$ bits) \cite{DBLP:journals/corr/CourbariauxB16},
$16$ bits are, as of today, the most frequently used bit-width. 
For the \acs{DNN} inference, this format is proven to be almost as accurate as single precision floating point weights \cite{farabet-ecvw-11}\cite{nn-X}\cite{Chen:2014:DSH:2541940.2541967}, whereas weight encodings with very few bits (e.g., $1$ or $2$ bits) suffer from comparable low accuracy \cite{DBLP:journals/corr/UmurogluFGBLJV16}.
Note that multiplications use $16$ bits, while the subsequent accumulation is done with $32$ bits.
This ensures that the input of the activation function is provided with a full precision of $32$ bits (e.g., Q15.16).

\subsection{Activation function}

All activation functions in our designs are implemented using comparators and arithmetic operations.
Due to the small number of cases in modern activation functions like \acs{ReLU}, an efficient implementation using combinational logic is possible while occupying only few logic resources.
More complex functions (e.g., sigmoid) are implemented using the piecewise linear approximation (\acs{PLAN}) that was originally proposed by Amin et al. \cite{PLAN}.
The desired function can be dynamically chosen by the \emph{AXI \acs{DNN} Control} which allows both accelerators to support different layer types.
Older implementations also used precomputed activation function images stored in lookup-tables \cite{Pietras6927383}.
However, as explained in \cite{PZ16} theses tables occupy valuable memory resources and are less flexible considering a dynamic change of the actual function type.

\subsection{Datapath Throughput Optimzation - Batch Processing}

\begin{figure}[t]
\centering
\includegraphics[width=0.9\linewidth]{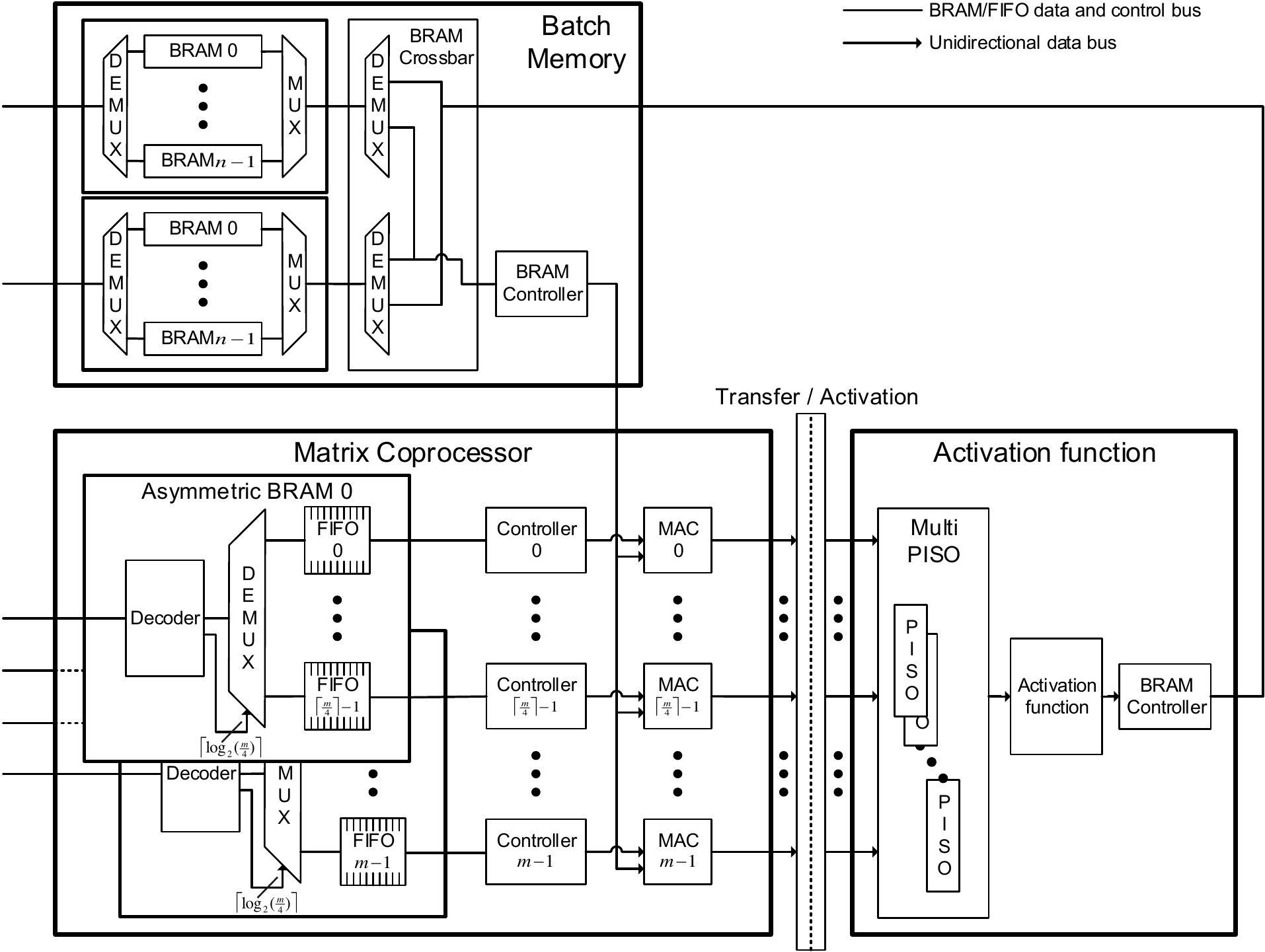}
\caption{Datapath for the batch processing of deep neural networks.
The batch memory contains two dedicated memory hierarchies for the previous and the current computed layer.
Each of the two memories contains $n$ storage elements for the $n$ processed samples.
The crossbar can switch between the input and output role of a layer.
The coprocessor and activation function implement the processing of $m \cdot n$ neurons before a software intervention is required.}
\label{fig:BatchDatapath} 
\end{figure}

In order to efficiently process multiple input samples, the previously discussed datapath components have to be adapted.
Figure~\ref{fig:BatchDatapath} shows the conceptual mapping of an arbitrary batch size with up to $n$ samples.

The memory hierarchy, here called \emph{Batch Memory}, contains $n$ \acsp{BRAM} for both input and output activations.
Due to the regular structure of the matrix-matrix operation in batch processing (cf. Section~\ref{sec:con:throughput}),
the \acs{BRAM} controller inside the batch memory can prefetch the correct input of the current section without stalling the pipeline and supply it to all $m$ parallel \acsp{MAC}. Note that in this architecture, all $m$ processing units, one for each neuron, have only a single MAC unit. Therefore, $r$ is set in this case to $r = 1$. 
At the \emph{same time}, the activations of the \emph{previous} section can be written into the memory.
The \acs{BRAM} crossbar facilitates that each \acs{BRAM} can play either the role of the input or the output, depending on the current processing state.

Using the batch memory hierarchy for the input activations and \acsp{FIFO} for the corresponding weights, the \emph{Matrix Coprocessor} calculates the transfer function for up to $m$ neurons in parallel.
The concrete number $m$ is only restricted by the number of available \acs{DSP} slices and \acs{BRAM} resources.
Often times the \acsp{BRAM} are the limiting factor for the number of parallel processing units since at least one \acs{FIFO} must be associated to one \acs{MAC} unit in order to supply the weights.
A \acs{FIFO} stores up to one row $w_{i,0}^{(j)} \dots w_{i,s_{j}-1}^{(j)}$ (the complete row if the previous layer is small enough) of the current weight matrix and is embedded in one of the four \emph{asymmetric \acsp{BRAM}} that are connected to the \acs{DMA} engines.

For the final computation of a neuron, the results of the coprocessor are passed to the activation function.
In case of batch processing, the complete design only requires one actual implementation of each function. 
A series of \emph{\acf{PISO}} registers is used to serialize the coprocessor outputs for a subsequent activation function.
A more detailed description can also be found in \cite{PZ16}.

Internally, all three datapath components of the batch processing design contain an extensive pipelining.
Although these pipeline stages exist, Figure~\ref{fig:BatchDatapath} visualizes only one pipeline register between the coprocessor and the activation function.
This stage is crucial for the batch processing since it allows a full decoupling of the transfer and activation function (i.e., both work in parallel using different samples).

Since we defined our section size to be $s_{j+1} \geq m$ and the coprocessor needs $s_{j}$ clock cycles for all \acs{MAC} operations of the section if $r =1$,
our activation function can take up to $s_j$ cycles before the next coprocessor results are in the main pipeline stage.
Furthermore, both the \acs{ReLU} and the sigmoid function are implemented using one clock cycle ($c_a = 1$).
Hence, our design of only one active activation function reduces the required \acs{FPGA} logic resources considerably without any throughput declines.
In general, the computation of the layer $j+1$ (including the cycles for the activation function) across all $n$ samples requires

\[
\Bigl\lceil \frac{s_{j+1}}{m} \Bigr\rceil \cdot s_j \cdot n + m \cdot c_a
\]


clock cycles since only the activations of the last section for sample $n$ are not computed in parallel ($m \cdot c_a$).
Moreover, for this term $m \cdot c_a \ll s_{j+1} \cdot s_j$ holds true.
Thus, the approximate time for calculating all results of layer $j+1$ is


\[
t_{calc} \approx \frac{ s_{j+1} \cdot s_j \cdot n}{m \cdot f_{pu}},
\]


which is the same as the formula in Section \ref{sec:con:throughput}  with $N = n$, $q_\text{prune}^{(j)} = 0$, and $r = 1$.

\subsection{Datapath Throughput Optimization - Pruning}
\label{sec:arch:pruning}

Compared to the batch processing design, where it is sufficient to transfer a sequence of weights and the dimension of the matrix operation, pruning requires additional metadata that gives information about the actual position of a weight $w_{i,k}^{(j)}$ within the matrix $W^{(j)}$ as stated in Section \ref{sec:con:pruning}.
We use a format similar to \cite{DBLP:journals/corr/HanLMPPHD16} that represents individual rows of the sparse weight matrices using tuples of $( w_l, z_{w_l})$ entries, with $l = 0 \dots (1 - q_{\text{prune},k}^{(j)}) \cdot s_j -1$.
Here, $w_l$ encodes a remaining weight after pruning and $z_{w_l}$ denotes the number of preceding zeros that come before $w_l$ in the corresponding row. 
The number of remaining weights after pruning is $s_j \cdot (1 - q_{\text{prune},k}^{(j)})$, where $q_{\text{prune},k}^{(j)}$ is the pruning factor of row $k$ of the weight matrix  $W^{(j)}$. The overall pruning factor $q_\text{prune}^{(j)}$ of the weight matrix  $W^{(j)}$ can be calculated with 
\[
q_\text{prune}^{(j)} = \frac{1}{s_{j+1}} \cdot \sum_{k = 0}^{s_{j+1} -1} q_{\text{prune},k}^{(j)}.
\]
Opposed to \cite{DBLP:journals/corr/HanLMPPHD16}, we do not separate the weights and zeros into two 1-dimensional arrays and store them in on-chip tables, but rather pack a certain number $r$ of consecutive $( w_l, z_{w_l})$ tuples into one \emph{data word} (cf. \cite{xapp1209}).
In our architecture we use $r = 3$ tuples, encode $w_l$ with the Q7.8 format (the same as in the batch processing approach), and represent $z_{w_l}$ as an unsigned integer with 5 bits.
Using these parameters, a row
\[
(0,\ -1.5,\ 0,\ 0,\ +0.3,\ -0.17,\ 0,\ 0,\ 0,\ +1.1,\ 0,\ 0,\ -0.2,\ 0,\ +0.1,\ \dots)
\]
is encoded into the following sequence of 64 bit data words
\vspace{0.2cm}
\begin{center}
\resizebox{0.9\linewidth}{!}{%
\begin{tikzpicture}[
    every node/.append style={draw,minimum width=0pt,minimum height=1.5em},
    >=latex,
    title/.style={font=\fontsize{8}{8}\color{black!50}\ttfamily,draw=none, inner sep=0.0cm, outer sep=0.0cm}
]

\tikzstyle{seperator}=[rectangle, node distance = 0.01mm and 0.01mm, minimum width = 0.01mm, draw=none]
\tikzstyle{element}=[rectangle, node distance = 1cm and 0.0cm, outer sep = 0.0cm]
\tikzstyle{padding}=[element, draw=gray!50, fill=gray!20]
\tikzstyle{packet}=[rectangle, node distance = 0cm and 0.0cm, outer sep = 0.0cm, inner sep=3pt]

\node[element, minimum width=1.6cm]                      (v1) {$-1.5$};
\node[element, minimum width=0.4cm, right of=v1]         (z1) {$1$};
\node[element, minimum width=1.6cm, right=.1 of z1.east] (v2) {$+0.3$};
\node[element, minimum width=0.4cm, right of=v2]         (z2) {$2$};
\node[element, minimum width=1.6cm, right=.1 of z2.east] (v3) {$-0.17$};
\node[element, minimum width=0.4cm, right of=v3]         (z3) {$0$};
\node[padding, minimum width=0.4cm, right=.1 of z3.east] (pad1) {};

\node[seperator, right=0.01 of pad1]  (sep1) {};

\node[element, minimum width=1.6cm, right of=sep1]       (v4) {$+1.1$};
\node[element, minimum width=0.4cm, right of=v4]         (z4) {$3$};
\node[element, minimum width=1.6cm, right=.1 of z4.east] (v5) {$-0.2$};
\node[element, minimum width=0.4cm, right of=v5]         (z5) {$2$};
\node[element, minimum width=1.6cm, right=.1 of z5.east] (v6) {$+0.1$};
\node[element, minimum width=0.4cm, right of=v6]         (z6) {$1$};
\node[padding, minimum width=0.4cm, right=.1 of z6.east] (pad2) {};

\node[seperator, right=0.01 of pad2]  (sep2) {};

\node[element, minimum width=1.6cm, right of=sep2, draw=none] (dots) {\dots};

\node (desc1) [title, above=0.1 of v1, anchor=west, xshift=-0.5cm, yshift=0.1cm] { data word $0$ };
\node (desc2) [title, above=0.1 of v4, anchor=west, xshift=-0.5cm, yshift=0.1cm] { data word $1$ };

\node[packet, fit={(v1)(z1)(v2)(z2)(v3)(z3)(desc1)(pad1)}]    (p1) {};
\node[packet, fit={(v4)(z4)(v5)(z5)(v6)(z6)(desc2)(pad2)}]    (p2) {};

\end{tikzpicture}%
}
\end{center}

Note that this encoding uses only 63 bit from the 64 bit available data word. The advantage is that the data is memory aligned to the 64 bit border which eases the memory access. The corresponding overhead per weight compared to non-pruning implementations is $q_\text{overhead} = 64 \text{ bit} / (3 \times 16 \text{ bit}) = 1.33$.

\begin{figure}[t]
\centering
\begin{overpic}[width=0.95\linewidth]{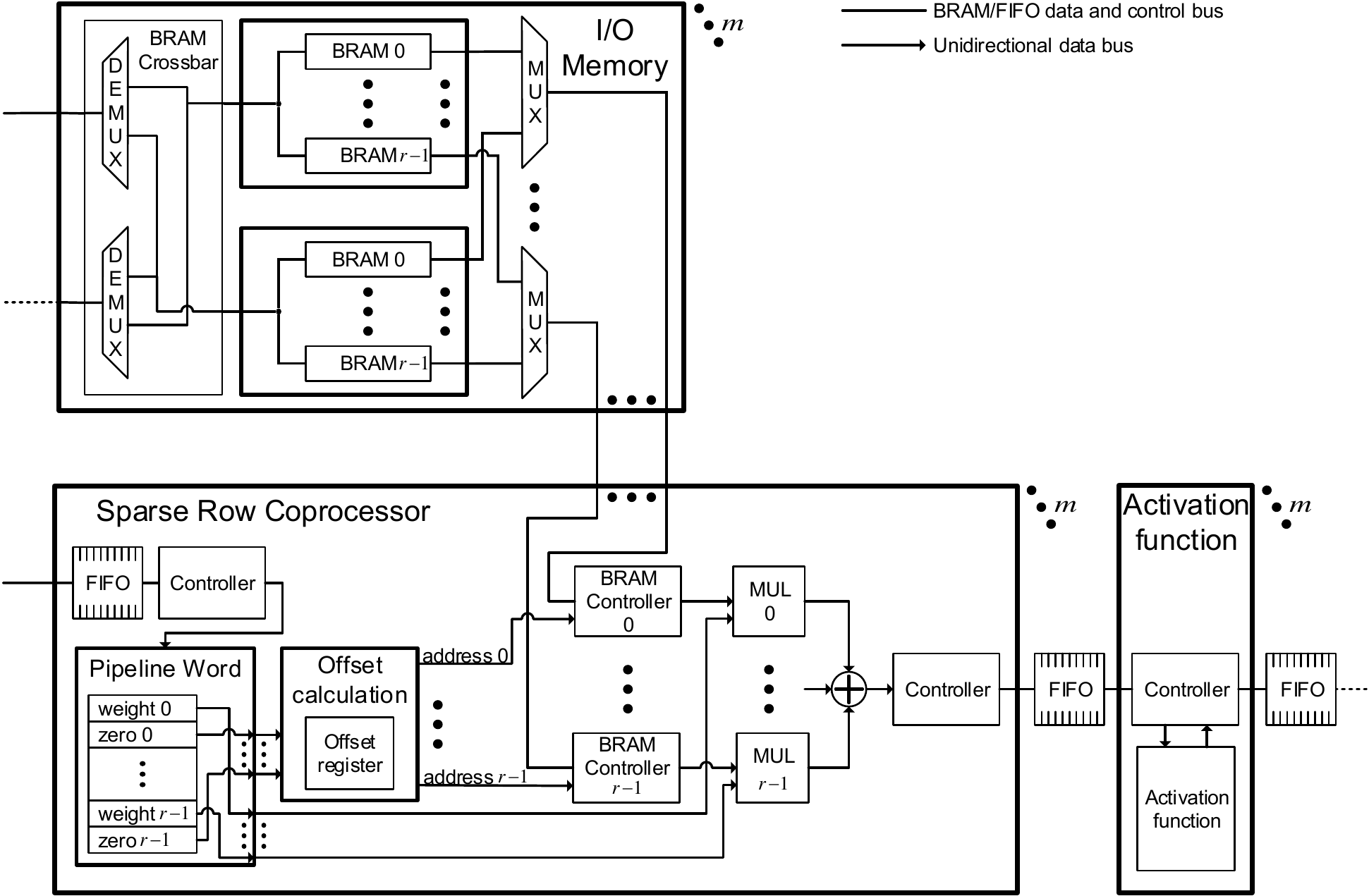}
  \put(2.75,-3.75){\includegraphics[width=0.95\linewidth, scale=0.95]{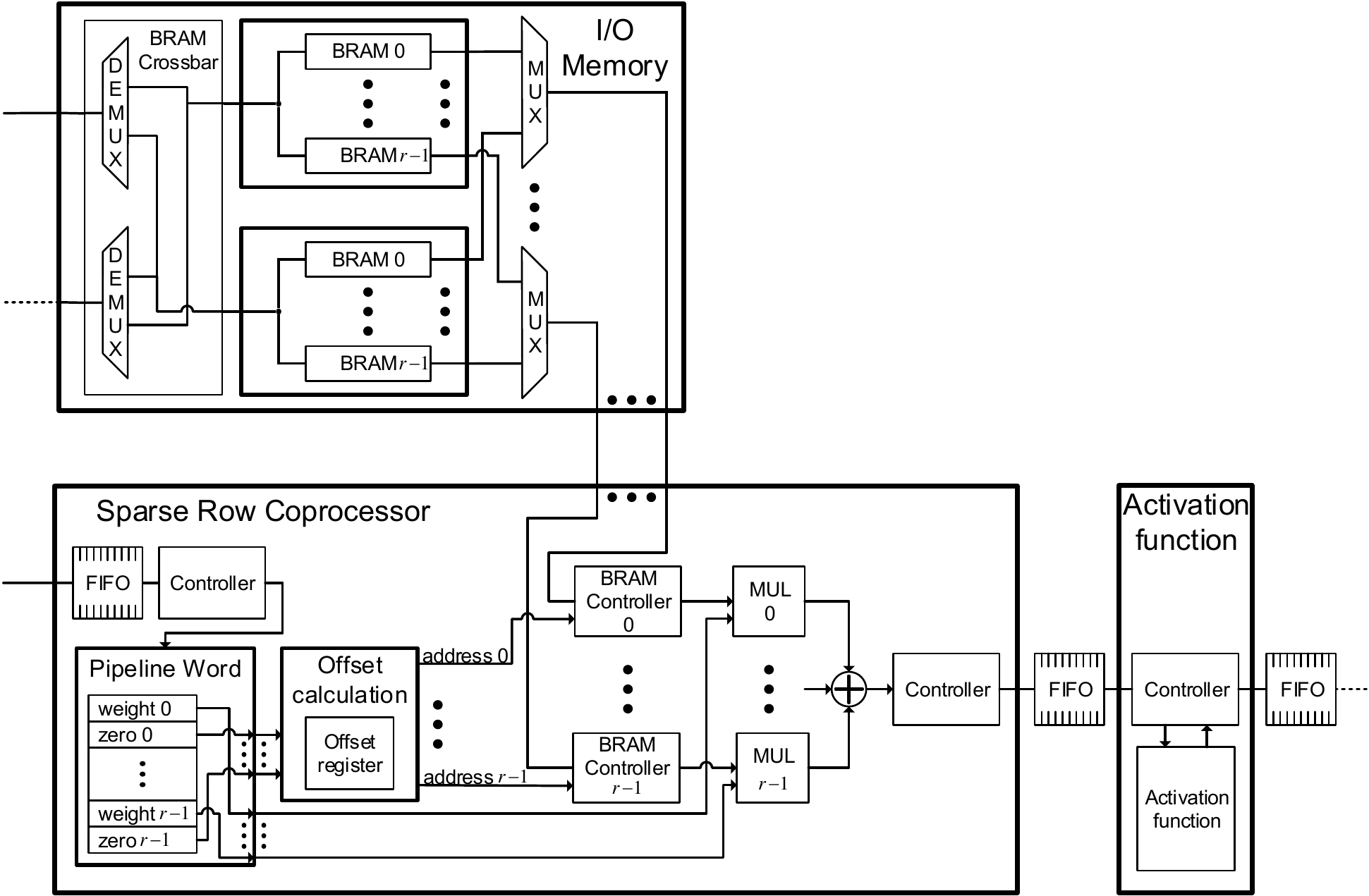}}
\end{overpic}
\vspace{0.4cm}
\caption{Datapath for the computation of sparse rows in pruned \acsp{DNN}.
This example presumes a pipeline word with $r$ tuples, each containing a weight and the number of zeros before it. 
In order to avoid delays when fetching the input activation that corresponds to a given weight, the \acsp{BRAM} in the I/O memory are also duplicated $r$ times, such that each multiplier has its own memory port.
By combining $m$ of these datapath instances, $m$ neurons can be computed in parallel (i.e., $m$ rows of the sparse matrix).
In such cases, an \acs{IP} that merges the activations of different rows must be connected with the I/O memories (indicated through the dashed lines).
}
\label{fig:PruningDatapath}
\end{figure}

Compared to other sparse matrix encodings that, for example, use separate vectors for the absolute row and column pointers \cite{vuduc2003automatic}, this format works well for streaming architectures since it directly combines both the weight and its relative position in one stream. This means that it does not require synchronization for, e.g., weight and multiple index streams.
Since the structure of pruned weight matrices is not as homogeneous as their dense counterparts,
the datapath of a corresponding streaming architecture must be designed to handle sparse matrices in order to avoid pipeline stalls.
A datapath adaptation that supports the discussed format is depicted in Figure~\ref{fig:PruningDatapath}.

Where the fully-connected structure assured that each input activation $a_k^{(j)}$ is needed for the computation of each neuron $a_i^{(j+1)}$ , in pruned weight matrices many input activations can be skipped due to corresponding zero-weights in the layer.
Hence, in the batch processing datapath an input activation $a_k^{(j)}$ is supplied to all $m$ parallel \acs{MAC} units, whereas in the pruning datapath the coprocessor needs to calculate the address of the input activation $a_k^{(j)}$ for the current weight.
This input address is potentially different for every row, which makes a parallel distribution of the inputs impractical. Therefore, each of the $m$ parallel sparse row coprocessors has it own I/O memory unit. This means that the I/O memory and the coprocessors are replicated $m$ times. 
Each of the $m$ I/O memories is addressed individually. To calculate the address in order to access the corresponding input activation  $a_k^{(j)}$, the following formula can be used:

\[
\text{address}_l =  l + \sum_{k = 0}^{l-1} z_{w_{k}}
\]

The \emph{offset calculation} \acs{IP} computes these addresses for all $r$ weights iteratively  using the previously computed and stored offset $o_\text{reg}$, the number of non-zero weights before $w_l$ and the zero fields $z_{w_{l}}$ from the pipeline word:
\[
\text{address}_i = o_\text{reg} + i + \sum_{k = 0}^{i} z_{w_{k}},
\]

where $i = 0 \dots r-1$. 
Depending on the number of tuples $r$, this means for the hardware design either a longer combinational path 
with $r$ subsequent adders or otherwise adders with up to $r+1$ inputs.
Since $r$ in our design is sufficiently small ($r = 3$), we use the latter.

Having computed the addresses, the coprocessor can multiply the weights and retrieve input activations and subsequently accumulate the partial sums.
However, in order to retrieve the weights in parallel and avoid multiple cycles for a sequential fetching of the individual activations, the input memory needs $r$ read ports.
Given that \acs{RAM} resources in current \acs{FPGA} technologies usually do not provide more than two memory ports \cite{blockram7series}, 
the \emph{I/O memory} inside the pruning datapath stores both input and output activations in $r$ redundant \acs{BRAM} copies.
This provides at any time the required $r$ memory ports.
Compared to the batch processing datapath, the I/O memories in the pruning datapath only store one sample.
When $m$ neurons should be computed in parallel, this redundancy is even increased to $m \cdot r$ copies since each of the $m$ coprocessors needs $r$ individual read ports. If the calculated $\text{address}_i$ surpasses the stored number of inputs $s_j$, the calculation of the current  transfer function $z_i^{(j+1)}$ is finalized, the result is handed over to the activation function, and the corresponding processing unit starts calculating the following transfer function $z_{i+m}^{(j+1)}$.
After the activation function, a merger \acs{IP} (not depicted in Figure~\ref{fig:PruningDatapath}) distributes the computed output activations of the $m$ neurons to \emph{all} I/O memories (second port of the \acs{BRAM} crossbar).
This requires a round-robin multiplexing scheme of the involved \acsp{FIFO} after the activation function.
Opposed to the batch processing datapath (cf. Figure~\ref{fig:BatchDatapath}), we decided to use $m$ hardware activation functions since the number of accumulations depends now on the percentage of the pruned parameters which might differ on a case-by-case basis.

\section{Experimental Results}
\label{sec:eval}

To evaluate and verify the so far discussed concepts, we have implemented both presented accelerators on an embedded platform and compared them with different configurations against miscellaneous software platforms.
In this section, we experimentally determine parameters like the best performing batch size $n$ and show how beforehand chosen parameters 
perform.
Furthermore, we show on the target hardware what performance gains are to be expected when both concepts are correspondingly implemented. 

We chose the \emph{Zynq Evaluation and Development Board} \cite{ZedBoardHWUserGuide}, short \emph{ZedBoard}, for the implementation of our designs.
It represents a typical embedded \acs{SoC} and features a XC7020 device with Artix-7 \acs{FPGA} fabric.

For the dedicated hardware support of different batch sizes we synthesized multiple bitstreams (a more detailed resource utilization is given in \cite{PZ16}) whereas the pruning design is only synthesized once with the parameters $m = 4$ and $r = 3$.

Each design uses two clock domains: the memory interface (e.g., Zynq high performance ports and \acsp{DMA}) is clocked with \mbox{133 MHz} and the remaining processing \acsp{IP} use a 100 MHz clock ($f_{pu}$).

\subsection{Throughput Evaluation}
\label{sec:perf}

For a fair comparison of both hardware and software, we have trained different fully-connected neural network architectures with multiple real-world data set.
As many before us, we use the famous \emph{\acs{MNIST} database of handwritten digits} \cite{MNIST} as the first benchmark.
The data set contains 60,000 training and 10,000 test samples.
A sample represents in this case a digit between $0$ and $9$, and is given as a grayscale image with a resolution of $28 \times 28$ pixels.
In addition, we have also performed all tests with a second benchmark that deals with the subject of recognizing human activities (\emph{\acs{HAR}}) of daily living through smartphone sensors \cite{HAR}.
For this purpose, a person (who is wearing the smartphone) performed one of six activities (walking, walking upstairs, walking downstairs, sitting, standing, and laying).
One sample of the data set is a 561-feature vector of time and frequency variables from different smartphone sensors (accelerometer, gyroscope, etc.).
A use case could be tracking of sport activities or, in a batch scenario, complete sequences of motions.
The data set is divided into 7,352 training and 2,947 test samples.

In our evaluation, all hardware candidates compete against a software implementation that we have tested on an embedded (i.e., the ZedBoard without FPGA use), a notebook (DELL Latitude E7250 Ultrabook) and, a desktop machine.
A more detailed hardware specification of all three platforms is given in Table~\ref{tbl:Competitors}.
Xilinx's bare-metal layer is used for the ZedBoard whereas both the notebook and the desktop machine use Linux-based operating systems.
By default, bare-metal uses only one core for the software execution.

\ctable[
pos = t,
caption = {Detailed hardware specification of the three machines used for the software \acs{DNN} processing.}, 
label = {tbl:Competitors},
doinside=\footnotesize,
] {p{3.2cm}|>{\centering\arraybackslash}p{1.8cm}|>{\centering\arraybackslash}p{1.8cm}|>{\centering\arraybackslash}p{1.8cm}} {
}
{
\FL
Machine &
\multirow{2}{*}[-0pt]{\minitab[c]{ARM \\ Cortex-A9}} & \multirow{2}{*}[-0pt]{\minitab[c]{Intel Core\\ i7-5600U}} & \multirow{2}{*}[-0pt]{\minitab[c]{Intel Core\\ i7-4790}}
\NN
&&&
\ML
CPU Clock Freq. (MHz) & 667 & 2600 - 3200 & 3600 - 4000
\ML
Cores (Threads) & 2 (2) 
 & 2 (4) & 4 (8)
\ML
L1 cache size (KB) & 32  & 128  & 256
\NN
L2 cache size (KB) & 512 & 512  & 1024
\NN
L3 cache size (KB) & --- & 4096 & 8192
\ML
Total \acs{RAM} (MB)& 512 & 8192 & 16384\NN
Dual channel used   &  no &  no  &  yes 
\ML
\acs{DDR}3 controller peak & 4.2~
& 12.8 & 25.6
\NN
bandwidth (GB/s) &&&
\LL
}

Furthermore, all presented processors feature some variant of a vector extension to accelerate floating-point intensive calculations through parallelism on instruction level.
For the ARM Cortex-A9 this extension is called NEON \cite{armneon} whereas both Intel CPUs can use SSE and AVX for this purpose \cite{firasta2008intel}.
In order to get the best runtime result on \emph{all} presented platforms, we use the \emph{\acs{BLAS}} \cite{OPENBLAS} library for the software inference of the \acsp{DNN}.
The library is individually configured and compiled for each of the used processors.
Note that the software is using 32-bit single-precision floating point numbers, whereas our hardware design uses the described Q7.8 fixed point format.
The throughput results for the \acs{DNN} inference on all software and hardware platforms are depicted in Table~\ref{tbl:PerformanceComparison}.

\ctable[
star,
pos = tbp,
caption = {Throughput comparison of our hardware-based batch processing (multiple configurations of hardware batch sizes), our hardware design with pruning support, and software inference on three different systems. Execution times are averaged over the size of the used test set and given in milliseconds (ms) per sample.},
label = {tbl:PerformanceComparison},
doinside=\footnotesize
] {l|l|c|c|c|c} {
\tnote[a]{ Network architectures: \; $784 \times 800 \times 800 \times 10$ \; and \; $784 \times 800 \times 800 \times 800 \times 800 \times 800 \times 800 \times 10$}
\tnote[b]{ Network architectures: \; $561 \times 1200 \times 300 \times 6$ \; and \; $561 \times 2000 \times 1500 \times 750 \times 300 \times 6$}
\tnote[c]{ Software calculations are performed using the \emph{IEEE 754 floating point single precision} format and using \emph{\acs{BLAS}}.
The i7-4790 utilizes dual channel memory whereas the others only use single channel.}
}
{
\multicolumn{2}{c}{} & \multicolumn{2}{|c|}{\acs{MNIST} \tmark[a]} & \multicolumn{2}{
c}{HAR \tmark[b]}
\NN
\FL
Device & Configuration & 4-layer netw. & 8-layer netw. & 4-layer netw. & 6-layer netw.
\NN
&& 1,275,200 & 3,835,200 & 1,035,000 & 5,473,800 
\NN
&& Parameters & Parameters & Parameters & Parameters
\ML
\multicolumn{6}{c}{Hardware-based batch processing} 
\ML
\textbf{Batch size 1} & 114 \acsp{MAC} & 1.543 & 4.496 & 1.3817 & 5.337
\ML
\textbf{Batch size 2} & 114 \acsp{MAC} & 0.881 & 2.520 & 0.7738 & 2.989
\ML
\textbf{Batch size 4} & 114 \acsp{MAC} & 0.540 & 1.505 & 0.463 & 1.792
\ML
\textbf{Batch size 8} & 106 \acsp{MAC} & 0.375 & 1.012 & 0.313 & 1.250
\ML
\rowcolor{VeryLightGray}
\textbf{Batch size 16} & ~90 \acsp{MAC} & 0.285 & 0.768 & 0.262 & 1.027
\ML
\textbf{Batch size 32} & ~58 \acsp{MAC} & 0.318 & 0.914 & 0.287 & 1.203
\ML
\multicolumn{6}{c}{Hardware-based pruning} 
\ML
\multicolumn{2}{r|}{Pruning factor} & 0.72 & 0.78 & 0.88 & 0.94
\ML
\textbf{Pruning design} & 12 \acsp{MAC} & 0.439 & 1.072 & 0.161 & 0.420
\ML
\multicolumn{6}{c}{Software-based processing\tmark[c]} 
\ML
\textbf{ARM}       & \#Threads: 1 & 16.151 & 48.603 & 13.120 & 70.240
\NN
\textbf{Cortex-A9} & & & & &
\ML
\textbf{Intel Core}& \#Threads: 1 & 0.285 & 1.603 & 0.223 & 2.246
\NN
\textbf{i7-5600U}  & \#Threads: 2 & 0.221 & 1.555 & 0.144 & 2.220
\NN
                   & \#Threads: 4 & 0.247 & 1.591 & 0.182 & 2.417
\ML
\textbf{Intel Core}& \#Threads: 1 & 0.118 & 0.917 & 0.114 & 1.406
\NN
\rowcolor{VeryLightGray}
\textbf{i7-4790}   & \#Threads: 4 & 0.057 & 0.569 & 0.045 & 1.205
\NN
                   & \#Threads: 8 & 0.065 & 0.687 & 0.055 & 1.491
\LL
}

In order to measure the inference times on the individual platforms, we query the hardware performance counters of the processors before and after the actual computation of the \acsp{DNN}.
Similarly, for our hardware accelerator, we use its control software (running on one ARM core) and read the cycle count before triggering the computation and after the computation is done.
In addition, the shown results are also averaged over the complete test set of the used benchmark.
The inference times (i.e., time difference) are then given in milliseconds (ms) per samples.

Besides different software and hardware platforms, we have also tested multiple neural network architectures which are taken or inspired from current research in the field.
For example, the smaller network for \acs{MNIST} was proposed in \cite{2015arXiv150302531H} while the larger one is an artificially extended version of that architecture with four additional hidden layers.

The best results for both hardware designs and all software runs are highlighted.
As visible, a pipeline with a batch size of 16 samples delivers the fastest ratio of input data and processing on the XC7020 target device.
The optimal calculated batch size $n_{opt}$ for the presented design is $12.66$, assuming a constant number of $m = 114$ processing units clocked with $f_{pu} = 100 \text{ MHz}$ and the used Q7.8 fixed point format. 
On the software side, we see the fastest inference for the desktop machine with a utilization of 4 threads and dual channel (\acs{DC}) memory.
The results of the ARM core are significantly slower than all other platforms.
A carefully written software implementation with \emph{fixed point} numbers (i.e., only 16 bits per weight) and the NEON extension could theoretically be about four times faster.
However, even then, the results would be multiple times slower than the hardware candidate with batch size 1 and more than an order of magnitude slower than most batch processing configurations.
On both the mobile and desktop CPU, the execution times depend mostly on the network size and, more precisely, on the matrix sizes of the individual layers.
While the matrices of both 4-layer networks fit completely into the CPU caches and thus enable a slightly faster execution times, the tables are turned for matrices of the deep learning era.
For example, the 6-layer \acs{HAR} network with a $2000 \times 1500$ matrix represents such a typical fully-connected layer.
Here, the hardware, despite its five times slower memory interface, clearly outperforms all software implementations.

As expected, the results of the pruning design are highly dependent on the actual pruning factor $q_\text{prune}$.
The \acs{MNIST} results with a pruning factor below 80\% are comparable to the performance of the batch processing design with batch size $n = 8$.
However, in the \acs{HAR} benchmark where more than 90\% of the parameters were pruned, the performance clearly surpasses the best batch processing results.
Due to the very limited amount of 4 high performance ports on the Zynq, our design utilizes only $m = 4$ coprocessors. This results in a total utilization of only $12$ \acsp{MAC}.

Furthermore, we compared our approach with a related FPGA-based neural network accelerator.
A fair and direct comparison is only possible with approaches that supply results for fully-connected \acsp{DNN} or \acsp{RNN} (\acsp{RNN} have only slightly more weights due to neuron feedback connections).
Apart from our presented batch processing scheme, accelerators for fully-connected layers can in general only use a weight once per network computation.
Instead, \acsp{CNN} are able to reuse weights due to a different neuron connection pattern in the convolutional layers.
Hence, they naturally achieve higher GOps/s due to a lower memory throughput requirement in direct comparisons with fully-connected layers.
However, when considering \emph{only} fully-connected layers the presented batch processing scheme clearly outperforms related work like, for example, a recent \acs{RNN} approach on the ZedBoard \cite{chang2015recurrent}.
The authors claim an overall throughput of 388.8 MOps/s.
With our approach and by using batch size $n = 16$, we reach a throughput of 4.48 GOps/s and 5.00 GOps/s, respectively (only counting \acs{MAC} operations). Although they are using less resources, our approach has a 6 times better throughput per DSP slice and 3 times better throughput per \acs{LUT} and \acs{FF}. The architecture using the pruning approach has only 0.8 GOps/s due to removed weights and operations.  However, compared with non-pruned approaches, this is equivalent to  2.91 GOps/s and 3.58  GOps/s, respectively. 

\subsection{Energy Efficiency}

Even though our approach outperforms almost all of the x86-based software configurations or has at least a comparable throughput, the real benefit is evident when comparing the energy efficiency. 
For determining the energy consumption, we measured the system power for processing the 8-layer neural network, introduced in Section~\ref{sec:perf}, and the idle power for all platforms (see Table~\ref{tbl:Energy}).
The overall power consumption on the ZedBoard is evaluated by measuring the average input voltage and the voltage drop on a shunt resistor. 
Whereas, the average power of the x86-based systems is measured on the primary side of the power supply with an ampere and volt meter.
Besides the idle and processing power, the energy consumption with (Overall Energy) and without (Dynamic Energy) the idle power consumption is shown in Table~\ref{tbl:Energy}.

\ctable[
pos = t,
caption = {Energy consumption comparison of our hardware designs and three processors (network: \acs{MNIST} 8-layer).},
label = {tbl:Energy},
doinside=\footnotesize
] {p{1.5cm}|l|c|c|c} {
}
{
\FL
Device & Configuration & Power & Overall & Dynamic
\NN
       &               & (W)   & Energy (mJ) & Energy (mJ)
\ML
\textbf{ZedBoard} & idle & 2.4  & --- & ---
\NN
\rowcolor{VeryLightGray}
 & HW batch ($n=16$)  & 4.4  & 3.8  & 1.5 
\NN
 & HW pruning ($m=4$) & 4.1  & 4.4  & 1.8 
\NN
 & SW \acs{BLAS} & 3.8  & 184.7  & 68.0 
\ML
\textbf{Intel} \textbf{Core} & idle & 8.9  & --- & ---
\NN
\rowcolor{VeryLightGray}
\textbf{i7-5600U}  & \#Threads: 1 & 20.7  & 33.2 & 18.9 
\NN
                   & \#Threads: 2 & 22.6  & 35.1 & 21.3 
\NN 
                   & \#Threads: 4 &  24.9 & 39.6 & 25.5 
\ML
\textbf{Intel} \textbf{Core} & idle & 41.4  & --- & ---
\NN
\textbf{i7-4790} & \#Threads: 1 & 65.8  & 63.9 & 22.4 
\NN
                 & \#Threads: 4 & 82.3  & 46.8 & 23.3 
\NN
                 & \#Threads: 8 & 81.8  & 56.2 & 27.8 
\LL
}

Comparing our best performing hardware configuration of batch size $n=16$ with pure software approaches, an overall energy efficiency improvement of almost factor 10 and more than factor 12 for the dynamic energy can be achieved.
In the latency measurements, the i7-5600U is the nearest competitor.

Compared to a competing LSTM design \cite{DBLP:journals/corr/HanKMHLLXLYWYD16}, our pruning approach is about factor 1.8 more energy efficient using their network with 3248128 weights, their pruning factor of $q_\text{prune} = 0.888$, and our theoretical throughput estimation of Section \ref{sec:con:throughput} (1.9 mJ for our pruning approach and 3.4 mJ for their approach). 

\subsection{Batch Processing Latency Evaluation}

\begin{figure}[t]
\centering
  \resizebox{0.75\linewidth}{!}
  {
  \begin{tikzpicture}
    \definecolor{grayblue}{rgb}{0.47, 0.66, 0.71}

    \begin{axis}[
      grid=both,
      scale only axis,
      xlabel = {Batch size},
      ylabel = {Average latency in ms},
      xtick  = {1,2,4,8,16,32},
      ymin   = 0,
      ymax   = 40,
      try min ticks=15,
      scaled ticks=false, tick label style={/pgf/number format/fixed},
      width  = 0.85\linewidth,
      height = 0.65\linewidth,
      every axis plot/.append style={thick},
      legend style={
        anchor=north east,
        at={(0.35,0.975)},
        cells={anchor=east},
        /tikz/every even column/.append style={column sep=0.5cm}
      },
    ]
      
      \addplot[Color1, mark=*]
      table[x index = 0,y index = 1, unbounded coords=discard]
      {Latency.txt};
      \addplot[Color2, mark=square*]
      table[x index = 0,y index = 2, unbounded coords=discard]
      {Latency.txt};
      \addplot[Color3, mark=diamond*]
      table[x index = 0,y index = 3, unbounded coords=discard]
      {Latency.txt};
      \addplot[Color4, mark=triangle*]
      table[x index = 0,y index = 4, unbounded coords=discard]
      {Latency.txt};
      
      \legend{MNIST 4-layer, MNIST 8-layer, HAR 4-layer, HAR 6-layer}
    \end{axis}
  \end{tikzpicture}
  }
\vspace{-2mm}
\caption{Latency analysis for different batch sizes and network architectures. Latency is given in milliseconds and averaged over the test set of the network.}
\label{fig:Latency}
\end{figure}
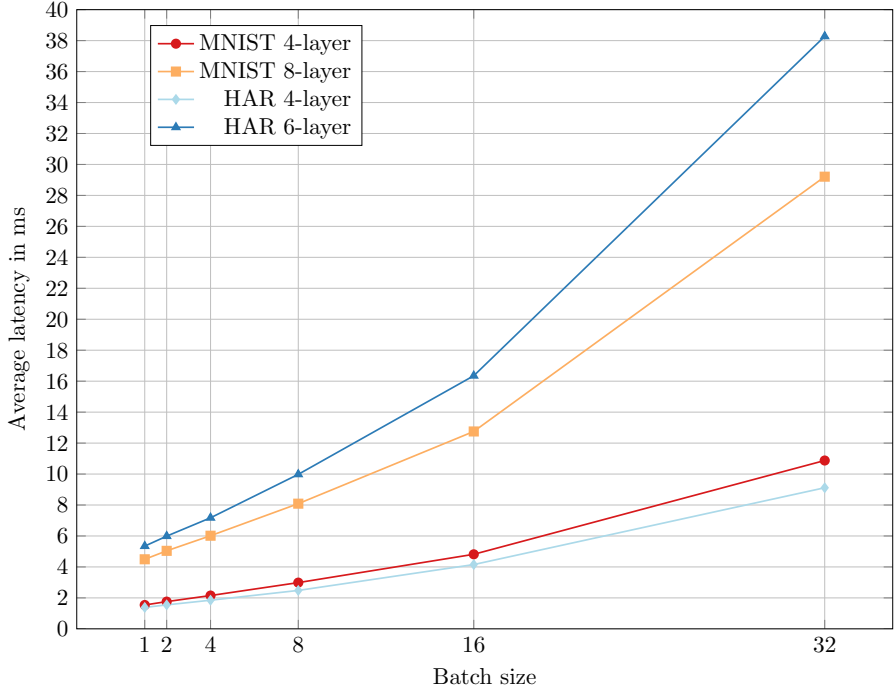

As mentioned earlier, the presented batch processing approach represents a trade-off between throughput and latency.
Figure~\ref{fig:Latency} compares the averaged latency of samples with the configured batch size.

For all of the investigated networks, a batch size of 8 samples results in approximately the doubled latency compared to regular processing with only 1 sample.
The best throughput configuration of batch size 16 yields approximately the tripled latency in comparison with regular processing.

\subsection{Accuracy Evaluation}

Since our batch processing accelerator utilizes all weights and the same Q7.8 fixed point data format as most related works \cite{farabet-ecvw-11}\cite{nn-X}\cite{10.1109/FCCM.2009.34}\cite{Chen:2014:DSH:2541940.2541967}, we obtain in this case similar results concerning the accuracy.
A more detailed description, that also takes a different ratio of integer and fractional bits into account, can be found in \cite{PZ16}.
The objective for the training with pruning was a maximum accuracy deviation of 1.5\% in correctly predicted samples.
All networks discussed in the throughput evaluation (i.e., Section~\ref{sec:perf}) meet this objective and deliver an accuracy very similar to their non-pruned counterparts (most deviate less than 0.5\%).
A detailed comparison of accuracy and pruning percentage is shown in Table~\ref{tbl:AccuracyPruning}.

\ctable[
star,
pos = t,
caption = {Accuracy evaluation in percentage of correctly predicted test set samples depending on the overall pruning factor $q_\text{prune}$ of the network.},
label = {tbl:AccuracyPruning},
mincapwidth = \textwidth,
doinside=\scriptsize
] {l|l|c|c|c|c} {
\tnote[a]{\tiny Network architectures: \; $784 \times 800 \times 800 \times 10$ \; and \; $784 \times 800 \times 800 \times 800 \times 800 \times 800 \times 800 \times 10$}
\tnote[b]{\tiny Network architectures: \; $561 \times 1200 \times 300 \times 6$ \; and \; $561 \times 2000 \times 1500 \times 750 \times 300 \times 6$}
}
{
\multicolumn{2}{c}{} & \multicolumn{2}{|c|}{\acs{MNIST} \tmark[a]} & \multicolumn{2}{
c}{HAR \tmark[b]}
\NN
\FL
\multicolumn{2}{r|}{Number of parameters} & 4-layer netw. & 8-layer netw. & 4-layer netw. & 6-layer netw.
\NN
\multicolumn{2}{r|}{} & 1,275,200 & 3,835,200 & 1,035,000 & 5,473,800 
\NN
\multicolumn{2}{r|}{} & Parameters & Parameters & Parameters & Parameters
\ML
\multicolumn{2}{r|}{Best non-pruned Accuracy} & \multicolumn{2}{c|}{98.3} & \multicolumn{2}{c}{95.9}
\ML
\multicolumn{2}{r|}{Pruning factor} & 0.72 & 0.78 & 0.88 & 0.94
\NN
\multicolumn{2}{r|}{Accuracy} & 98.27 & 97.62 & 94.14 & 95.72
\ML
}

\newpage

\section{Conclusions and Future Work}
\label{sec:conc}

In this paper, we present two architectures for an \acsp{FPGA}-based embedded SoC that are able to accelerate the inference of previously learned fully-connected deep neural networks.
Both designs mitigate a slow access to the external memory on such SoCs by either reusing weights across multiple input samples (batch processing) or by pruning weights that do not affect the accuracy.
Our comparison shows that these orthogonal techniques are both able to substantially increase the throughput of \acs{DNN} accelerators.
However, while batch processing does not affect the \acs{DNN} accuracy, it may only be used in scenarios where an increased latency is tolerable.
On the contrary, pruning can possibly reduce the accuracy but also even surpass the batch processing throughput results.
In general, we were able to prune at least over 70\% of parameters without noticeable accuracy drops and, for example, process $8$ input samples in parallel with just a doubled latency.
Additionally, each presented technique outperforms fully-featured x86-based systems once the size of the weight matrices is larger than the available cache.
Thereby, each technique results in an energy-efficiency that is more than an order of magnitude better.
Similarly, while large \acsp{FPGA} outperform our design in terms of pure GOps/s, our design implemented on the presented embedded \acs{FPGA} is almost factor two more energy-efficient.

Future works on this topic might further increase the throughput by combining both techniques into one datapath.
The theoretical results, calculated from the formulas in the throughput discussion in Section \ref{sec:con:throughput}, show that such a combination would substantially increase the throughput. However, one problem might be the used memory resources. Both approaches need a high amount of additional on-chip memories which are scarce on small embedded devices. Nevertheless, an envisaged design with $m = 6$, $r = 3$, and $n = 3$ would be feasible on the used Zynq and would have an expected inference time of the 6-layer HAR of 186 $\mu s$. This would be over 6 times faster than our fastest used x86 processor system.


\newcommand{\acro}{\acrodef}
  \acro{ANN}{Artificial Neural Network}
  \acro{DNN}{Deep Neural Network}
  \acro{CNN}{Convolutional Neural Network}
  \acro{RNN}{Recurrent Neural Network}
  \acro{ReLU}{Rectified Linear Unit}
  \acro{PLAN}{Piecewise Linear Approximation of a Nonlinear Function}
  \acro{FPGA}{Field Programmable Gate Array}
  \acro{TDM}{Time Division Multiplexing}
  \acro{GPU}{Graphics Processing Unit}
  \acro{ASIC}{Application Specific Integrated Circuit}
  \acro{NPU}{Neural Processing Unit}
  \acro{PT}{Processing Tile}
  \acro{MAC}{Multiply Accumulate}
  \acro{BRAM}{Block \acs{RAM}}
  \acro{FIFO}{First In, First Out}
  \acro{PISO}{Parallel In, Serial Out}
  \acro{SoC}{System on a Chip}
  \acro{PS}{Processing System}
  \acro{PL}{Programmable Logic}
  \acro{IP}{Intellectual Property}
  \acro{DSP}{Digital Signal Processing}
  \acro{LUT}{Lookup Table}
  \acro{FF}{Flip-Flop}
  \acro{RAM}{Random Access Memory}
  \acro{DDR}{Double Data Rate}
  \acro{DMA}{Direct Memory Access}
  \acro{OCR}{Optical Character Recognition}
  \acro{MNIST}{Mixed National Institute of Standards and Technology}
  \acro{HAR}{Human Activity Recognition}
  \acro{BLAS}{Basic Linear Algebra Subprograms}
  \acro{SC}{Single Channel}
  \acro{DC}{Double Channel}

\bibliography{bib}

\end{document}